\documentclass[12pt]{article}

\font\grande=cmr9.5 scaled \magstep4
\font\medio=cmr9.5 scaled \magstep2
\outer\def\beginsection#1\par{\medbreak\bigskip
      \message{#1}\leftline{\bf#1}\nobreak\medskip
\vskip-\parskip
      \noindent}
\usepackage{graphicx} 
\begin{document}
\bibliographystyle {unsrt}

\titlepage

\begin{flushright}
CERN-PH-TH/2010-067
\end{flushright}

\vspace{10mm}
\begin{center}
{\grande A circular polarimeter}\\
\vspace{5mm}
{\grande for the Cosmic Microwave Background}\\
\vspace{1.5cm}
 Massimo Giovannini 
 \footnote{Electronic address: massimo.giovannini@cern.ch} \\
\vspace{1cm}
{{\sl Department of Physics, 
Theory Division, CERN, 1211 Geneva 23, Switzerland }}\\
\vspace{0.5cm}
{{\sl INFN, Section of Milan-Bicocca, 20126 Milan, Italy}}
\vspace*{0.5cm}
\end{center}

\vskip 1cm
\centerline{\medio  Abstract}
A primordial degree of circular polarization of the Cosmic Microwave Background is not observationally excluded.  The hypothesis of primordial dichroism can be 
quantitatively falsified if  the plasma is magnetized prior to photon decoupling since 
the initial V-mode polarization affects the evolution of the temperature fluctuations as well as the equations for the linear polarization. The observed values of the temperature and polarization angular power spectra are used to infer constraints on the amplitude and on the spectral slope of the primordial V-mode.  Prior to photon decoupling magnetic fields play the role of polarimeters insofar as they unveil the circular dichroism by coupling the V-mode power spectrum to the remaining brightness perturbations.  Conversely, for angular scales ranging between 4 deg and 10 deg the joined bounds on the magnitude of circular polarization and on the magnetic field intensity suggest that direct limits on the V-mode power spectrum in the range of $0.01$ mK could directly rule out pre-decoupling magnetic fields in the range of $10$--$100$ nG.  The frequency dependence of the signal is located, for the present purposes, in the GHz range. 
\noindent

\vspace{5mm}

\vfill
\newpage
\renewcommand{\theequation}{1.\arabic{equation}}
\setcounter{equation}{0}
\section{Introduction}
\label{sec1}
The electron-photon scattering in a magnetized plasma provides computable source terms for the evolution of the brightness perturbations \cite{pera} which can be studied with diverse initial conditions, under various kinds of approximations and in different physical systems  ranging from the classic problem of line formation of a normal Zeeman triplet \cite{unno} to sychrotron emission \cite{sync1,sync2,sync3,sync4}. 
In conventional Cosmic Microwave Background (CMB) studies the initial conditions of the temperature and polarization anisotropies are provided by the standard adiabatic mode \cite{bert1,bert2}. The latter 
requirement implies that the initial radiation field lacks a specific degree of linear polarization. 
By initial radiation field we simply mean, in the present context, 
the initial data for the four Stokes parameters prior to matter-radiation equality.  

In the adiabatic scenario, the collision terms of the brightness perturbations for electron-photon 
scattering (see, e.g. \cite{pera}) allow for the generation of linear polarization after photon decoupling 
while the circular polarization is, comparatively, not affected, i.e. it is vanishing both prior to matter-radiation equality 
and after photon decoupling. The study of the linear polarization 
(and of its potential generation via electron-photon scattering) has a long history
\cite{h1,h2,h3,h4,h5} dating even before the hypothesis of adiabatic initial conditions 
triggered by the formulation of inflationary scenarios. 
The preliminary detection of the polarization autocorrelations \cite{QUAD1,QUAD2} (i.e. the EE power spectrum) as well as of the temperature-polarization correlations \cite{WMAP7a,WMAP7b,WMAP7c,WMAP7d,WMAP7e,WMAP7f} (i.e. the TE power spectrum) confirms that the initial data, prior to matter-radiation equality, are predominantly adiabatic and lacking any specific degree of linear polarization which arises, to leading order in the tight-coupling expansion, because of the quadrupole of the temperature fluctuations. 

In a series of recent papers \cite{mg1,mg2}, the idea has been to assume that the initial data of the radiation field are dictated by the conventional adiabatic mode but that the electron-photon scattering takes place in a magnetized environment.  An amount of circular polarization is then produced because of the properties of magnetized electron-photon scattering.  Analogies with the latter phenomenon  can be found in the physics of the magnetized sunspots as well as in complementary astrophysical situations (see, for instance, \cite{TS1,TS2}). The circular polarization arising from unpolarized initial conditions has then been computed and, to lowest order in the tight-coupling expansion, is proportional to the monopole of the intensity of the radiation field \cite{mg1,mg2} (see also \cite{BHR} for a different kind of derivation).  

The situation explored in \cite{mg1,mg2} does not exhaust the possible sets of initial data for the system of brightness 
perturbations. The statement that the initial radiation 
field does not possess a specified (linear) polarization does not forbid, however, the presence of a primordial (circular) polarization. If the pre-decoupling plasma is not magnetized, then, the circular polarization will evolve independently  both from the temperature fluctuations as well as from the linear polarization. Conversely, 
if the plasma is magnetized, the circular polarization can directly affect both the temperature anisotropies as well as the E-mode polarization.  

In the present paper we want to address a situation which is opposite to the one 
investigated in \cite{mg1,mg2}. The idea will be to assume 
that the initial radiation field has an unknown amount of circular polarization. 
The purpose  will then be to constrain the primordial V-mode by using the magnetized plasma as a polarimeter 
rather than as a polarizer. In \cite{mg1,mg2} the magnetic field acted as a polarizer: 
the circular polarization was assumed to vanish initially 
and the problem was to compute the resulting V-mode signal. In the present paper  the magnetic field will act as a polarimeter: the initial circular polarization does not vanish and the presence of the magnetic field is used as a diagnostic on the initial radiation field. 

The layout of the paper will therefore be the following. 
In section \ref{sec2} the relevant governing equations will be introduced. 
In section \ref{sec3} the V-mode contribution to the 
temperature autocorrelations (TT correlations for short) will be computed 
and constrained. In section \ref{sec4} the same exercise will be repeated 
in the case of the polarization autocorrelations (EE correlations for short).
The bounds obtained in sections \ref{sec3} and \ref{sec4} will be 
used in section \ref{sec5} with the aim of inferring constraints  on the autocorrelations 
of the circular polarization (VV power spectra for short). Section \ref{sec6} contains the concluding remarks and the perspectives for future studies. 

\renewcommand{\theequation}{2.\arabic{equation}}
\setcounter{equation}{0}
\section{Brightness perturbations in magnetized plasmas}
\label{sec2}
The results on the evolution of brightness perturbations 
in magnetized plasmas will be briefly recapitulated and the essential notations 
introduced. The derivation of the equations can be found in \cite{mg1,mg2} and will not be repeated here. The evolution equations for the brightness perturbations can be written, in Fourier space,  as\footnote{The evolution 
of brightness perturbations appearing in Eqs. (\ref{deltaI}), (\ref{deltaP}) and 
(\ref{deltaV}) is discussed in the conformally Newtonian gauge 
where the perturbed entries of the geometry read 
$\delta_{\mathrm{s}} g_{00} = 2 a^2 \phi$ and $\delta_{\mathrm{s}} g_{ij} = 2 a^2 \psi 
\delta_{ij}$. It should be noticed that the conventions are 
different from Ref. \cite{bert1,bert2} where $\phi$ and $\psi$ are exchanged and where the signature of the metric is  mostly minus.}:
\begin{eqnarray}
\Delta_{\mathrm{I}}' + ( i k \mu + \epsilon') \Delta_{\mathrm{I}} &=& \psi' - i k\mu \phi + \epsilon'\biggl[ \Delta_{\mathrm{I}0} + 
\mu v_{\mathrm{b}} - \frac{P_{2}(\mu)}{2} S_{\mathrm{P}}\biggr] 
\nonumber\\
&-& \frac{3}{2} i\, \epsilon' \, f_{\mathrm{e}}(\omega) \,( 1 + \mu^2) \Delta_{\mathrm{V}1} 
\label{deltaI}\\
 \Delta_{\mathrm{P}}' + ( i k \mu + \epsilon') \Delta_{\mathrm{P}} &=& \frac{3}{4} ( 1 - \mu^2) \epsilon' S_{\mathrm{P}}  - \frac{3}{2} i \epsilon' f_{\mathrm{e}}(\omega) (\mu^2 -1) \Delta_{\mathrm{V}1},
\label{deltaP}\\
  \Delta_{\mathrm{V}}' + ( i k \mu + \epsilon') \Delta_{\mathrm{V}}  &=& \epsilon' \mu \biggl\{ f_{\mathrm{e}}(\omega) [ 2 \Delta_{\mathrm{I}0} - S_{\mathrm{P}}] - \frac{3}{4} i \Delta_{\mathrm{V}1}\biggr\},
\label{deltaV}
\end{eqnarray}
where the prime denotes a derivation with respect to $\tau$, i.e. the conformal time coordinate\footnote{In this paper conformally flat geometries will be considered 
where the background metric  
 $\overline{g}_{\mu\nu} = a^2(\tau)$ is written in terms of the Minkowski metric 
 $\eta_{\mu\nu}= \mathrm{diag}(+1,\, -1,\, -1, \, -1)$ and of the scale factor $a(\tau)$. This choice 
 is directly dictated by the adopted set of fiducial parameters (see below Eqs. (\ref{parameters1}) and (\ref{parameters2})).}. Moreover, $S_{\mathrm{P}}$, $f_{\mathrm{e}}(\omega)$ and $\epsilon'$ are defined as 
\begin{eqnarray}
&& S_{\mathrm{P}} = \Delta_{\mathrm{P}2} + \Delta_{\mathrm{P}0} + \Delta_{\mathrm{I}2},\qquad f_{\mathrm{e}}(\omega) = \frac{\omega_{\mathrm{Be}}}{\omega},
\label{def1}\\
&& \epsilon' = x_{\mathrm{e}} \,\tilde{n}_{\mathrm{e}}\, \sigma_{\gamma\mathrm{e}}\, \frac{a}{a_{0}}, \qquad \epsilon(\tau,\tau_{0}) = \int_{\tau}^{\tau_{0}} x_{\mathrm{e}} \,\tilde{n}_{\mathrm{e}}\, \sigma_{\gamma\mathrm{e}}\, \frac{a}{a_{0}} \, d\tau.
\label{def2}
\end{eqnarray}
Concerning Eqs. (\ref{def1}) and (\ref{def2}) few comments are in order. The limit $f_{\mathrm{e}}(\omega) \to 0$ corresponds to the standard situation where 
the plasma is not magnetized: indeed $f_{\mathrm{e}}(\omega)$ denotes the ratio between the Larmor frequency of the electrons and the angular frequency of the 
observational channel. As   usual $\epsilon'$ is the differential optical depth while $\epsilon(\tau,\tau_{0})$ is the optical depth.
It is appropriate to remind the explicit form of $f_{\mathrm{e}}(\omega)$ which does depend upon the angular frequency $\omega = 2 \pi \nu$ and upon the magnetic field intensity $B_{\mathrm{u}}$
\begin{equation}
f_{\mathrm{e}}(\omega) = \frac{\omega_{\mathrm{Be}}}{\omega}
 = 2.79 \times 10^{-12} \biggl(\frac{B_{\mathrm{u}}}{\mathrm{nG}}\biggr)
 \biggl(\frac{\mathrm{GHz}}{\nu}\biggr) (z_{*} +1),\qquad \omega_{\mathrm{Be}}=
 \frac{e  |\vec{B}\cdot\hat{n}|}{m_{\mathrm{e}}a},
 \label{FE}
 \end{equation}
where $z_{*}$ is the redshift to last scattering, i.e. $z_{*} = 1090.79_{-0.92}^{+0.94}$ according to the WMAP-7yr data.  In the present paper the values of the cosmological 
parameters will be taken to coincide, without loss of generality, with the $\Lambda$CDM 
best fit to the WMAP 7-yr data alone; the critical fractions of baryons, CDM particles and 
dark energy will be, given respectively, by 
\begin{equation}
 ( \Omega_{\mathrm{b}}, \, \Omega_{\mathrm{c}}, \Omega_{\mathrm{de}}) = (0.0449\pm0.0028, 0.222\pm 0.026, 0.734\pm 0.029),
\label{parameters1}
\end{equation}
while the rescaled Hubble parameter, the adiabatic spectral index and the optical 
depth to reionization are  
\begin{equation}
(h_{0},\,n_{\mathrm{s}},\, \epsilon_{\mathrm{re}}) =(0.710\pm 0.025, 0.963\pm 0.014, 0.088 \pm 0.015).
\label{parameters2}
\end{equation}
The power spectrum of the curvature perturbations will be assigned, as usual, at the pivot scale $k_{\mathrm{p}} = 0.002\, \mathrm{Mpc}^{-1}$ as 
\begin{equation}
{\mathcal P}_{{\mathcal R}}(k) = {\mathcal A}_{{\mathcal R}} \biggl(\frac{k}{k_{\mathrm{p}}}\biggr)^{n_{\mathrm{s}} -1}, \qquad {\mathcal A}_{{\mathcal R}} = (2.43 \pm 0.11) \times 10^{-9}.
\label{parameters3}
\end{equation}
The system of equations given in (\ref{deltaI}), (\ref{deltaP}) and (\ref{deltaV}) can be studied in different physical limits. If we assume that 
the initial V-mode polarization vanishes exactly we can keep, at least initially 
$\Delta_{\mathrm{V}1}=0$ and Eqs. (\ref{deltaI})--(\ref{deltaV}) reduce to 
\begin{eqnarray}
\Delta_{\mathrm{I}}' + ( i k \mu + \epsilon') \Delta_{\mathrm{I}} &=& \psi' - i k\mu \phi + \epsilon'\biggl[ \Delta_{\mathrm{I}0} + 
\mu v_{\mathrm{b}} - \frac{P_{2}(\mu)}{2} S_{\mathrm{P}}\biggr] , 
\label{deltaI1}\\
 \Delta_{\mathrm{P}}' + ( i k \mu + \epsilon') \Delta_{\mathrm{P}} &=& \frac{3}{4} ( 1 - \mu^2) \epsilon' S_{\mathrm{P}},
 \label{deltaP1}\\
 \Delta_{\mathrm{V}}' + ( i k \mu + \epsilon') \Delta_{\mathrm{V}}  &=& \epsilon' \mu f_{\mathrm{e}}(\omega) [ 2 \Delta_{\mathrm{I}0} - S_{\mathrm{P}}].
\label{deltaV1}
\end{eqnarray}
The system of Eqs. (\ref{deltaI1})--(\ref{deltaV1}) describes the situation where 
the radiation field does not have (initially)  any circular polarization. As a result 
of the electron-photon scattering in a magnetized environnment a tiny amount 
of circular polarization is produced according to Eq. (\ref{deltaV1}) and has been 
computed in \cite{mg1,mg2}. In the present paper the logic will be opposite to the one leading to Eqs. (\ref{deltaI1})--(\ref{deltaV1}). 
More specifically, it will be assumed that the initial radiation 
field {\em is}  circularly polarized (i.e. $\Delta_{\mathrm{V}1} \neq 0$ initially).
The question will then be how large could  $\Delta_{\mathrm{V}1}$ not to 
affect the temperature and polarization autocorrelations.  
In \cite{mg2} the coupling to $\Delta_{\mathrm{V}1}$ appearing in the 
analog of Eqs. (\ref{deltaP}) and (\ref{deltaV}) has been incorrectly squared. 
We take here the chance of correcting this coupling which does not affect the results 
of \cite{mg2} since, in that context, $\Delta_{\mathrm{V}1}$ did vanish prior to matter radiation equality. 

The maximum of the microwave background arises today for typical photon energies of the order of $10^{-3}$ 
eV corresponding to a typical wavelength of the mm. At the time of photon decoupling (i.e. $z\simeq z_{*} \simeq 1090$)  the wavelength of the radiation was of the order 
of $10^{-3} \mathrm{mm} \simeq \mu\mathrm{m}$.  Since the magnetic field 
we are interested in is inhomogeneous on a scale comparable with the Hubble radius 
(at the corresponding epoch)  the guiding 
centre approximation \cite{gcapp} can be safely employed as already discussed 
in \cite{mg1,mg2}. There are different ways of introducing the guiding centre approximation and the simplest one is to think of a gradient expansion 
of the background magnetic field, i.e. denoting  with $\vec{B}$ the (comoving) 
magnetic field intensity we can write that
\begin{equation}
B_{i}(\vec{x},\tau) \simeq B_{i}(\vec{x}_{0}, \tau)  + (x^{j} - x_{0}^{j}) \partial_{j} B_{i} +...
\label{co6a}
\end{equation}
where the ellipses stand for the higher orders in the gradients leading, both, to curvature and drift corrections which will be neglected in this investigation as they were 
neglected in \cite{mg1,mg2}. Higher order in the gradients demand the 
inclusion of higher multipoles of the field (see e. g. \cite{beckers}). The latter 
effects can be neglected to lowest order in the guiding centre approximation. 

\renewcommand{\theequation}{3.\arabic{equation}}
\setcounter{equation}{0}
\section{Temperature autocorrelations}
\label{sec3}
The temperature fluctuations can be written as 
\begin{equation}
\Delta_{\mathrm{T}}(\hat{n},\tau_{0}) = \sum_{\ell m} a^{(\mathrm{T})}_{\ell m} \, 
Y_{\ell m}(\hat{n}), \qquad \hat{n}=(\vartheta, \, \varphi)
\label{TT1}
\end{equation}
From the line of sight solution of Eq. (\ref{deltaI}), 
the power spectrum of the temperature correlations receives two 
separated contributions stemming, respectively, from the intensity 
of the radiation field (denoted as  $\overline{a}^{(\mathrm{I})}_{\ell \, m}$) 
and from the circular polarization (denoted as  $\overline{a}^{(\mathrm{V})}_{\ell \, m}$):
\begin{equation}
a^{(\mathrm{T})}_{\ell\, m} = \overline{a}^{(\mathrm{I})}_{\ell \, m} + \overline{a}^{(\mathrm{V})}_{\ell \, m}.
\label{a1}
\end{equation}
To make the notations clear we want to stress that 
$\overline{a}^{(\mathrm{V})}_{\ell \, m}$ denotes the V-mode contribution to 
the temperature correlation and not the power spectrum of the V-mode itself 
(which will be introduced later on in section \ref{sec5}).
According to Eq. (\ref{deltaI}) the two terms at the right hand side can be written as 
\begin{eqnarray}
&&  \overline{a}^{(\mathrm{I})}_{\ell \, m} = \frac{1}{(2\pi)^{3/2}} \int d^{3} k \int_{-1}^{1} \, d\mu \, \int_{0}^{2\pi} d\varphi  \,Y_{\ell\, m}^{*}(\mu,\varphi) \int_{0}^{\tau_{0}} 
\, e^{- i \mu x} \, e^{- \epsilon(\tau,\tau_{0})} \, {\mathcal N}_{\mathrm{I}}(k,\mu,\tau) \, d\tau,
\label{a2}\\
&&  \overline{a}^{(\mathrm{V})}_{\ell \, m} = \frac{1}{(2\pi)^{3/2}} \int d^{3} k \int_{-1}^{1} \, d\mu \, \int_{0}^{2\pi} d\varphi  \,Y_{\ell\, m}^{*}(\mu,\varphi) \int_{0}^{\tau_{0}} 
\, e^{- i \mu x} \, e^{- \epsilon(\tau,\tau_{0})} \, {\mathcal N}_{\mathrm{V}}(k,\mu,\tau) \, d\tau,
\label{a3}
\end{eqnarray}
where $\mu = \cos{\vartheta}$; the two generalized sources ${\mathcal N}_{\mathrm{I}}(k,\mu,\tau)$ and ${\mathcal N}_{\mathrm{V}}(k,\mu,\tau)$ are given, respectively,  by:
\begin{eqnarray}
&& {\mathcal N}_{\mathrm{I}}(k,\mu,\tau) = \psi' - i k \mu \phi + \epsilon' \biggl[ \Delta_{\mathrm{I}0} + \mu v_{\mathrm{b}} - \frac{1}{2} P_{2}(\mu)S_{\mathrm{P}}\biggr],
\label{a4}\\
&& {\mathcal N}_{\mathrm{V}}(k,\mu,\tau) = - \frac{3}{2} \, i\, \epsilon' f_{\mathrm{e}}(\omega) ( 1 + \mu^2) 
\Delta_{\mathrm{V}1}.
\label{a5}
\end{eqnarray}
For  angular scales larger than 1 deg, the visibility function can be approximated with a sufficiently thin \footnote{Note that $\tau_{0}$ denotes the angular diameter distance to $\tau_{*}$. It is appropriate to mention 
that the visibility function has also a second (smaller) peak which arises because the Universe is reionized at late times.  The reionization peak affects the overall amplitude 
of the CMB anisotropies and polarization.  It also affects the peak structure 
of the linear polarization. The fiducial set of parameters 
adopted in Eqs. (\ref{parameters1})--(\ref{parameters2}) 
 suggest that the typical redshift for reionization 
is $z_{\mathrm{reion}} =10.5 \pm 1.2 $ and the corresponding optical depth is $\epsilon_{\mathrm{re}}= 0.088\pm 0.015$. For semi-analytic purposes the second peak can be modeled with a second Gaussian profile \cite{zalpol}. } Gaussian profile \cite{zeld1,pav1,wyse}:
\begin{equation} 
{\mathcal K}(\tau) = \epsilon' \, e^{- \epsilon(\tau,\tau_{0})} = \sqrt{\frac{2}{\pi}} \frac{1}{\sigma_{*}} e^{- \frac{(\tau- \tau_{*})^2}{2 \sigma_{*}^2}},
\label{a6}
\end{equation} 
i.e. $\tau_{0} \gg \tau_{*}$ and $\tau_{0} \gg \sigma_{*}$.  In the latter limit  and assuming that the V-mode and the adiabatic contribution are uncorrelated the angular power spectrum at large scales is given by
\begin{eqnarray}
C_{\ell}^{(\mathrm{TT})} = {\mathcal F}_{\ell}^{(\mathrm{I})} + {\mathcal F}_{\ell}^{(\mathrm{V})},
\label{a7}
\end{eqnarray}
where the two terms at the right hand side are, from Eq. (\ref{a1}),
\begin{equation}
 {\mathcal F}_{\ell}^{(\mathrm{I})} = \frac{1}{2\ell +1} \sum_{m = - \ell}^{\ell}  \langle 
  |\overline{a}^{(\mathrm{I})}_{\ell \, m}|^2\rangle, \qquad 
  {\mathcal F}_{\ell}^{(\mathrm{V})} = \frac{1}{2\ell +1} \sum_{m = - \ell}^{\ell}  \langle 
 |\overline{a}^{(\mathrm{V})}_{\ell \, m}|^2\rangle.
\label{a8}
\end{equation}
According to Eq. (\ref{a7}), the temperature autocorrelations contain two physically 
different terms: the first contribution is given by the fluctuations 
of the intensity of the radiation field; the second contribution stems from the V-mode. The second contribution vanishes either when the magnetic field is zero or in the case when the circular polarization of primordial origin vanishes. To derive Eqs. (\ref{a7}) and (\ref{a8}) it has been assumed that the V-mode contribution is not correlated with the adiabatic mode which is instead responsible for the perturbation in the intensity of the radiation field \cite{bert1,bert2}.

It is now useful to derive the evolution equations for the lowest multipoles of the V-mode polarization. Recalling the present conventions for the Rayleigh expansion 
\begin{eqnarray}
&& \Delta_{\mathrm{I}}(k,\mu,\tau) = \sum_{\ell} \, (-i)^{\ell} \, (2 \ell +1) P_{\ell}(\mu) \, \Delta_{\mathrm{I}\ell}(k,\tau),
\label{r1}\\
&& \Delta_{\mathrm{V}}(k,\mu,\tau) = \sum_{\ell} \, (-i)^{\ell} \, (2 \ell +1) P_{\ell}(\mu) \, \Delta_{\mathrm{V}\ell}(k,\tau),
\label{r2}
\end{eqnarray}
the following set of relations can be easily derived  for the monopole, 
for the dipole and for the higher multipoles of the brightness perturbation associated with the V-mode:
\begin{eqnarray}
&& \Delta_{\mathrm{V}0}' + \epsilon' \Delta_{\mathrm{V}0} + k \Delta_{\mathrm{V}1} =0,
\label{mon1}\\
&& \Delta_{\mathrm{V}1}' + \frac{2}{3}k \Delta_{\mathrm{V}2} - \frac{k}{3} \Delta_{\mathrm{V}0} = 
- \frac{3}{4} \epsilon' \Delta_{\mathrm{V}1},
\label{dip1}\\
&& \Delta_{\mathrm{V}\ell}' + \epsilon' \Delta_{\mathrm{V}\ell} 
= \frac{k}{2 \ell + 1} [ \ell \Delta_{\mathrm{V}(\ell-1)} - (\ell + 1) \Delta_{\mathrm{V}(\ell + 1)}].
\label{mult}
\end{eqnarray}
In analogy with what customarily done in the case of the brightness perturbations 
of the intensity it is practical to define 
\begin{eqnarray}
&&\delta_{\mathrm{V}} = \frac{1}{\pi} \int d\mu\, d\varphi \Delta_{\mathrm{V}}(k,\mu,\tau) 
= 4 \, \Delta_{\mathrm{V}0}, 
\label{int0}\\
&&\theta_{\mathrm{V}} = \frac{3i}{4\pi} \int d\mu\, d\varphi \mu\, \Delta_{\mathrm{V}}(k,\mu,\tau)  
= 3 \, k \, \Delta_{\mathrm{V}1},
\label{int1}\\
&& \sigma_{\mathrm{V}} = -\frac{3}{4\pi }\int d\mu\, d\varphi 
\biggl( \mu^2 - \frac{1}{3}\biggr) \Delta_{\mathrm{V}}(k,\mu,\tau) =  2\Delta_{\mathrm{V}2} .
\label{def2a}
\end{eqnarray}
In terms of $\delta_{\mathrm{V}}$, $\theta_{\mathrm{V}}$ and $\sigma_{\mathrm{V}}$ Eqs. (\ref{mon1}) and (\ref{dip1}) can be written, over large scales, as 
\begin{eqnarray}
&& \delta_{\mathrm{V}}' + \frac{4}{3} \theta_{\mathrm{V}} =0, 
\label{delV}\\
&&\theta_{\mathrm{V}}' + k^2 \sigma_{\mathrm{V}} = \frac{k^2}{4} 
\delta_{\mathrm{V}}.
\label{thetaV}
\end{eqnarray}
It is therefore possible 
to set initial conditions by requiring that, prior to matter-radiation equality, the power 
spectrum of the dipole does not vanish and it is given as  
\begin{equation}
{\mathcal P}_{\mathrm{V}}(k) = {\mathcal A}_{\mathrm{V}} \biggl(\frac{k}{k_{\mathrm{p}}}\biggr)^{n_{\mathrm{v}} -1}, \qquad k_{\mathrm{p}} = 0.002\, \mathrm{Mpc}^{-1},
\label{PV}
\end{equation}
where $k_{\mathrm{p}}$ is the same pivot scale used to assign the adiabatic mode 
in Eq. (\ref{parameters3}). By combining Eqs. (\ref{mon1}) and (\ref{dip1}) the evolution of the  dipole can be 
written as:
\begin{equation}
\theta_{\mathrm{V}}'' + \frac{k^2}{3} \theta_{\mathrm{V}} =  - k^2 \sigma_{\mathrm{V}}.
\label{comb}
\end{equation}
Setting to zero all the multipoles $\ell \geq 2$, the solution of Eq. (\ref{comb}) prior to equality will be given by:
\begin{equation}
\theta_{\mathrm{V}}(k,\tau) = C_{1}(k) \cos{(c_{\mathrm{s}v} k \tau)} + C_{2}(k) \sin{(c_{\mathrm{s}v} k \tau)},
\label{comb2}
\end{equation}
 where $c_{\mathrm{s}v} \simeq 1/\sqrt{3}$. Equation (\ref{comb2}) stipulates that there are two separate classes of initial conditions: initial conditions where, prior to equality, the degree of circular polarization vanishes (i.e. $C_{1}(k) =0$ implying that $\theta(k,\tau) \to 0$ for $\tau\to 0$); initial conditions where, prior to equality, the degree of circular polarization goes to a constant (i.e. $C_{2}(k) =0$ implying that 
  $\theta(k,\tau) \to C_{1}(k)$ for $\tau\to 0$).  

The purpose of the present investigation is to set a bound on the amount of 
circular polarization present prior to equality and it is therefore appropriate to postulate that $C_{2}(k) =0$ while the constant $C_{1}(k)$  is directly related to the power spectrum of Eq. (\ref{PV}).
More specifically we shall have, for $\tau \ll \tau_{\mathrm{eq}}$, that 
 $|\Delta_{\mathrm{V}1}(k,\tau)|^2 =  2\pi^2\,{\mathcal P}_{\mathrm{V}}(k)/k^3$; however, because of Eq. 
 (\ref{int1}), and in the same limit, the following chain of equalities holds 
 \begin{equation}
 |\theta_{\mathrm{V}}(k,\tau)|^2 = |C_{1}(k)|^2 = 
 9 k^2 |\Delta_{\mathrm{V}1}(k)|^2 = \frac{18 \pi^2}{k} {\mathcal P}_{\mathrm{V}}(k).
 \label{comb3}
 \end{equation}
For large angular scales, the sudden decoupling limit implies that the angular power spectrum of the circular polarization is given by:
\begin{eqnarray}
 {\mathcal F}_{\ell}^{(\mathrm{V})} &=& \frac{9 \pi^2}{2} f_{\mathrm{e}}^2(\omega) \biggl(\frac{k_{0}}{k_{\mathrm{p}}}\biggr)^{n_{\mathrm{v}}-1}\,\, {\mathcal A}_{\mathrm{V}} \,\, {\mathcal I}_{\mathrm{V}}(\ell,n_{v}),
\label{CellV1}\\
{\mathcal I}_{\mathrm{V}}(\ell,n_{v}) 
&=& A^2(\ell)  {\mathcal I}_{1}(\ell,n_{v}) + B^2(\ell)  {\mathcal I}_{2}(\ell,n_{v}) + 
C^2(\ell)  {\mathcal I}_{3}(\ell,n_{v}) 
\nonumber\\
&-& 2 A(\ell) B(\ell) {\mathcal I}_{4}(\ell, n_{v}) - 2 A(\ell) C(\ell) {\mathcal I}_{5}(\ell, n_{v}) 
\nonumber\\
&+&  2 C(\ell) B(\ell) {\mathcal I}_{6}(\ell, n_{v}). 
\label{CellV2}
\end{eqnarray}
The coefficients $A(\ell)$, $B(\ell)$ and $C(\ell)$ appearing in Eq. (\ref{CellV2}) are the same as the ones 
obtained in Eq. (\ref{AA8}) of appendix \ref{APPA}. The integrals
 ${\mathcal I}_{i}(\ell,n_{v})$ (with $i= 1,\,...6$) can be computed 
 analytically and the result is
\begin{eqnarray}
{\mathcal I}_{1}(\ell,n_{v}) &=& \frac{1}{2 \sqrt{\pi}} \frac{\Gamma\biggl(\frac{3}{2} - \frac{n_{v}}{2} \biggr)
 \Gamma\biggl(\ell + \frac{n_{v}}{2} - \frac{1}{2}\biggr)}{\Gamma\biggl(2 - \frac{n_{v}}{2}\biggr) \Gamma\biggl(\frac{5}{2} + \ell - \frac{n_{v}}{2}\biggr)},
\label{CellV4}\\
 {\mathcal I}_{2}(\ell,n_{v}) &=& \frac{1}{2 \sqrt{\pi}} \frac{\Gamma\biggl(\frac{3}{2} - \frac{n_{v}}{2} \biggr)
 \Gamma\biggl(\ell + \frac{n_{v}}{2} - \frac{5}{2}\biggr)}{\Gamma\biggl(2 - \frac{n_{v}}{2}\biggr) \Gamma\biggl(\frac{1}{2} + \ell - \frac{n_{v}}{2}\biggr)},
 \label{CellV5}\\
{\mathcal I}_{3}(\ell,n_{v}) &=& \frac{1}{2 \sqrt{\pi}} \frac{\Gamma\biggl(\frac{3}{2} - \frac{n_{v}}{2} \biggr)
 \Gamma\biggl(\ell + \frac{n_{v}}{2} + \frac{3}{2}\biggr)}{\Gamma\biggl(2 - \frac{n_{v}}{2}\biggr) \Gamma\biggl(\frac{9}{2} + \ell - \frac{n_{v}}{2}\biggr)},
 \label{CellV6}\\
 {\mathcal I}_{4}(\ell,n_{v}) &=& - \frac{1}{4 \sqrt{\pi}} \frac{(n_{v} -2)\Gamma\biggl(\frac{3}{2} - \frac{n_{v}}{2}\biggr) \Gamma\biggl(\ell - \frac{3}{2} + \frac{n_{v}}{2} \biggr)}{\Gamma\biggl(3 - \frac{n_{v}}{2}\biggr) 
 \Gamma\biggl(\frac{3}{2} +\ell - \frac{n_{v}}{2} \biggr)},
 \label{CellV7}\\
 {\mathcal I}_{5}(\ell,n_{v}) &=& - \frac{1}{4 \sqrt{\pi}} \frac{(n_{v} -2)\Gamma\biggl(\frac{3}{2} - \frac{n_{v}}{2}\biggr) \Gamma\biggl(\ell + \frac{1}{2} + \frac{n_{v}}{2} \biggr)}{\Gamma\biggl(3 - \frac{n_{v}}{2}\biggr) 
 \Gamma\biggl(\frac{7}{2} +\ell - \frac{n_{v}}{2} \biggr)},
 \label{CellV8}\\
 {\mathcal I}_{6}(\ell,n_{v}) &=& \frac{2}{\sqrt{\pi}} \frac{\Gamma\biggl(\frac{3}{2} - \frac{n_{v}}{2}\biggr) 
 \Gamma\biggl(\frac{n_{v}}{2} + \ell - \frac{1}{2}\biggr)}{(n_{v} - 6) (n_{v}- 4) 
 \Gamma\biggl(\frac{5}{2} + \ell - \frac{n_{v}}{2}\biggr) 
 \Gamma\biggl(- \frac{n_{v}}{2}\biggr)},
 \label{CellV9}
 \end{eqnarray}
 where we assumed $-3 < n_{v} < 3$. By demanding that the intensity 
power spectrum is always smaller than the corresponding V-mode 
contribution, Eqs. (\ref{a7}) and (\ref{a8}) imply that
\begin{equation}
{\mathcal F}_{\ell}^{(\mathrm{V})} < {\mathcal F}_{\ell}^{(\mathrm{I})} \simeq C_{\ell}^{(\mathrm{TT})}.
\label{boundTT}
\end{equation}
Equation (\ref{boundTT}) holds in the standard $\Lambda$CDM scenario and under the assumption that the V-mode contribution and the adiabatic contribution are not correlated. The inclusion of the cross correlation may entail 
the addition of a further power spectrum and will not be explicitly discussed in this paper.
From Eq. (\ref{a2}), after integrating once by parts the term $- ik \mu \phi$,
it is easy to obtain the following expression for $\overline{a}^{(\mathrm{I})}_{\ell\, m}$
\begin{equation}
\overline{a}^{(\mathrm{I})}_{\ell\, m} = \frac{(-i)^{\ell} \delta_{m \, 0}}{(2\pi)^{3/2}} \,\sqrt{ 2 \ell + 1} \,\sqrt{4 \pi}
\int d^{3} k\, \Delta_{\mathrm{I}\ell}(k,\tau_{0}).
\label{boundTT1}
\end{equation}
The term $\Delta_{\mathrm{I}\ell}(k,\tau_{0})$ can be read off directly from Eq. (\ref{r1}) and 
by bearing in mind the line of sight solution of the heat transfer equation. The result is
\begin{equation}
\Delta_{\mathrm{I}\ell}(k,\tau_{0})  =
\biggl[ \Delta_{\mathrm{I}0}(k, \tau_{*}) + \phi(k,\tau_{*})\biggr] j_{\ell}[ k (\tau_{0} - \tau_{*})]
+ \int_{\tau_{*}}^{\tau_{0}}\, (\phi' + \psi') \, j_{\ell}[ k (\tau_{0} - \tau)]\, d\tau,
\label{boundTT2}
\end{equation}
where $j_{\ell}(x)$ are the spherical Bessel functions \cite{abr1,abr2};  in Eq. 
(\ref{boundTT2}) the first term corresponds to the ordinary Sachs-Wolfe contribution while the second term corresponds 
to the integrated Sachs-Wolfe effect which operates, in the $\Lambda$CDM scenario, after matter-radiation 
equality.  Over sufficiently large angular scales the difference between $\phi$ and $\psi$ can be ignored and Eq. (\ref{boundTT1}) 
can be written as:
\begin{equation}
\Delta_{\mathrm{I}\ell}(k,\tau_{0})  = - \frac{{\mathcal R}_{*}(k)}{5} \, j_{\ell}[k (\tau_{0} - \tau_{*})]
- 2 \int_{\tau_{*}}^{\tau_{0}} \biggl(\frac{d T}{d\tau}\biggr) \, {\mathcal R}_{*}(k) \, j_{\ell}[ k (\tau_{0} - \tau)] d\tau,
\label{boundTT3}
\end{equation}
where 
\begin{equation}
T(\tau) = 1 - \frac{{\mathcal H}}{a^2} \int_{0}^{\tau} a^2(\tau') d\tau',
\label{boundTT4}
\end{equation}
as it can be easily deduced, by direct integration with respect to the conformal 
time coordinate, from the definition of curvature perturbations in the 
conformally Newtonian gauge:
\begin{equation}
{\mathcal R} = - \psi - \frac{{\mathcal H} ({\mathcal H} \phi + \psi')}{{\mathcal H}^2 - {\mathcal H}'},
\label{bounTT4a}
\end{equation}
and under the assumption, already mentioned,  that $\phi=\psi$.
It is appropriate to recall that the first thorough analytical discussion 
of the integrated Sachs-Wolfe effect in the $\Lambda$CDM paradigm 
dates back to the results of Ref. \cite{alstar}.
The integral over the conformal time appearing in Eq. (\ref{boundTT3}) can be performed by recalling the 
the first zero of the spherical Bessel function is given  by $k(\tau_{0} - \tau_{k}) \simeq  (2\ell +1)/2$, i.e.
$\tau_{k} = \tau_{0} - (2\ell +1)/(2 k)$. Thus we can also write that 
\begin{equation}
\int_{\tau_{*}}^{\tau_{0}} \biggl(\frac{d T}{d\tau}\biggr) \, {\mathcal R}_{*}(k) \, j_{\ell}[ k (\tau_{0} - \tau)] d\tau
\simeq \frac{1}{k} \biggl(\frac{d T}{d\tau}\biggr)_{\tau= \tau_{k}}{\mathcal R}_{*}(k) \, \int_{0}^{\infty} j_{\ell}(x) \,d x.
\label{boundTT5}
\end{equation}
Equation (\ref{boundTT3}) becomes then:
\begin{equation}
\Delta_{\mathrm{I}\ell}(k,\tau_{0})  = - \frac{{\mathcal R}_{*}(k)}{5} \, j_{\ell}[k (\tau_{0} - \tau_{*})]
- \biggl(\frac{d T}{d\tau}\biggr)_{\tau= \tau_{k}}{\mathcal R}_{*}(k) 
\frac{\sqrt{\pi}}{ k} 
\frac{\Gamma\biggl(\frac{\ell +1}{2}\biggr)}{\Gamma\biggl(\frac{\ell + 2 }{2}\biggr)}.
\label{boundTT6}
\end{equation}
The exact solution of the Einstein equations in a $\Lambda$CDM cosmology can be 
written, in the cosmic time coordinate, as
\begin{equation}
a(t) = a_{1} \biggl\{
\sinh{\biggl[\frac{3}{2} \sqrt{\Omega_{\Lambda0}} H_{0} (t -t_0)\biggr]}\biggr\}^{2/3}, \qquad a_{1} =  \biggl(\frac{\Omega_{\mathrm{M}0}}{\Omega_{\Lambda0}}\biggr)^{1/3}.
\label{boundTT7}
\end{equation}
The conformal time coordinate $\tau(\alpha)$ and $T(\alpha)$ (the transfer function)  can then be expressed in terms of integrals over the (normalized) scale factor 
$\alpha$ as:
\begin{equation}
\tau(\alpha) = \frac{2}{a_{1} H_{0} \sqrt{\Omega_{\Lambda 0}}} \int^{\sqrt{\alpha}} \frac{ dy }{\sqrt{y^{6}+1}},
\qquad 
T(\alpha) = 1 - \frac{\sqrt{\alpha^3 +1}}{\alpha^{3/2}} \int_{0}^{\alpha} \frac{\beta^{3/2}}{\sqrt{\beta^3 +1}}\, d\beta.
\label{boundTT9}
\end{equation}
The two integrals of Eq. (\ref{boundTT9}) 
can be performed analytically in terms of elliptic functions; however the obtained result must anyway be integrated (numerically) over  $k$. This will mean that, at the end we will not have a closed analytic expression. We will probably gain in accuracy 
but the simplicity of the results will be partially lost. It is useful to notice, in this respect, that the ordinary and the integrated Sachs-Wolfe contributions are reasonably well separated in scales.  
 The Sachs-Wolfe contribution typically 
peaks for comoving wavenumbers $k \simeq 0.0002 \, \mathrm{Mpc}^{-1}$ while the integrated  Sachs-Wolfe effect contributes between $k_{\mathrm{min}} =0.001\, \mathrm{Mpc}^{-1}$ and $k_{\mathrm{max}}= 0.01\, \mathrm{Mpc}^{-1}$. For comparison recall that the pivot scale $k_{\mathrm{p}} =0.002\,\,
\mathrm{Mpc}^{-1}$ corresponds, for the best fit parameters of the WMAP 7yr alone (and in the light of the $\Lambda$CDM scenario to $\ell_{\mathrm{p}} \simeq 30$. The latter remark suggest that if the integrated 
Sachs-Wolfe contribution is neglected, the ordinary Sachs-Wolfe term leads to a more constraining bound simply because it does not always dominate at large scales.  The strategy will then  be to derive an analytic form of the bound by considering only the ordinary Sachs-Wolfe contribution.  More refined treatments could certainly improve on this aspect but, for the present purposes, the estimate will be sufficiently accurate as we shall see in a moment. By inserting Eq. (\ref{boundTT6}) into Eq. (\ref{boundTT1}) the angular power spectrum becomes 
\begin{eqnarray}
{\mathcal F}^{(\mathrm{I})}_{\ell} &=& \frac{2\pi^2}{25} \biggl(\frac{k_{0}}{k_{\mathrm{p}}}\biggr)^{n_{s} -1} 
{\mathcal A}_{{\mathcal R}} e^{- 2\, \epsilon_{\mathrm{re}}}\, {\mathcal I}_{{\mathcal R}}(\ell, n_{s}),
\label{boundTT10}\\
 {\mathcal I}_{{\mathcal R}}(\ell, n_{s}) &=& \frac{1}{2\sqrt{\pi}} \frac{\Gamma\biggl(\frac{3}{2} - \frac{n_{s}}{2}\biggr) \Gamma\biggl(\ell + \frac{n_{s}}{2} - \frac{1}{2}\biggr)}{\Gamma\biggl(2 - \frac{n_{s}}{2} \biggr) \Gamma\biggl(\ell + \frac{5}{2} -\frac{n_{s}}{2} \biggr)},
\label{boundTT11}
\end{eqnarray}
where $\epsilon_{\mathrm{re}}$ accounts for the overall suppression 
inherited from the reionization peak of the visibility function \cite{zalpol}.
\begin{figure}[!ht]
\centering
\includegraphics[height=6cm]{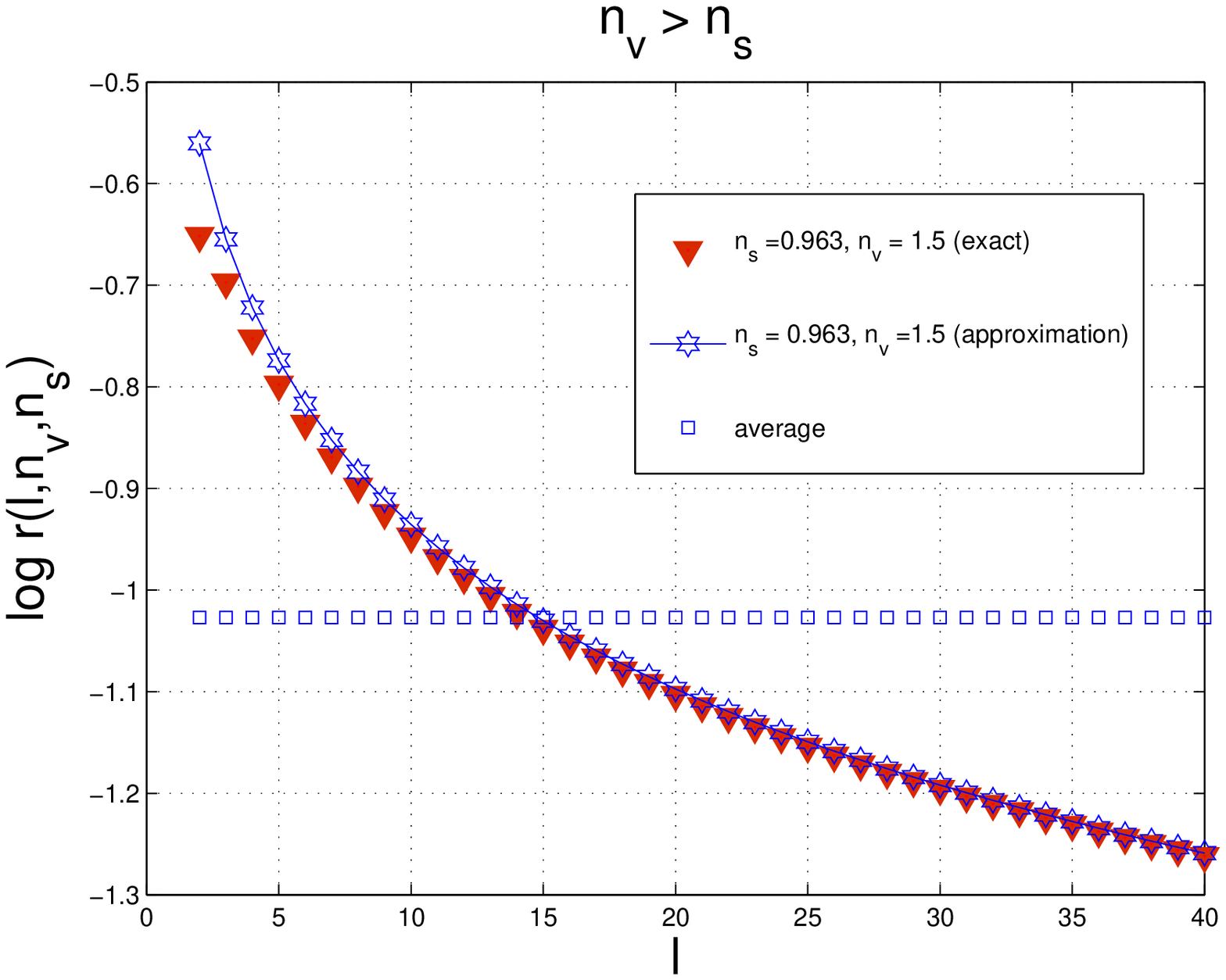}
\includegraphics[height=6cm]{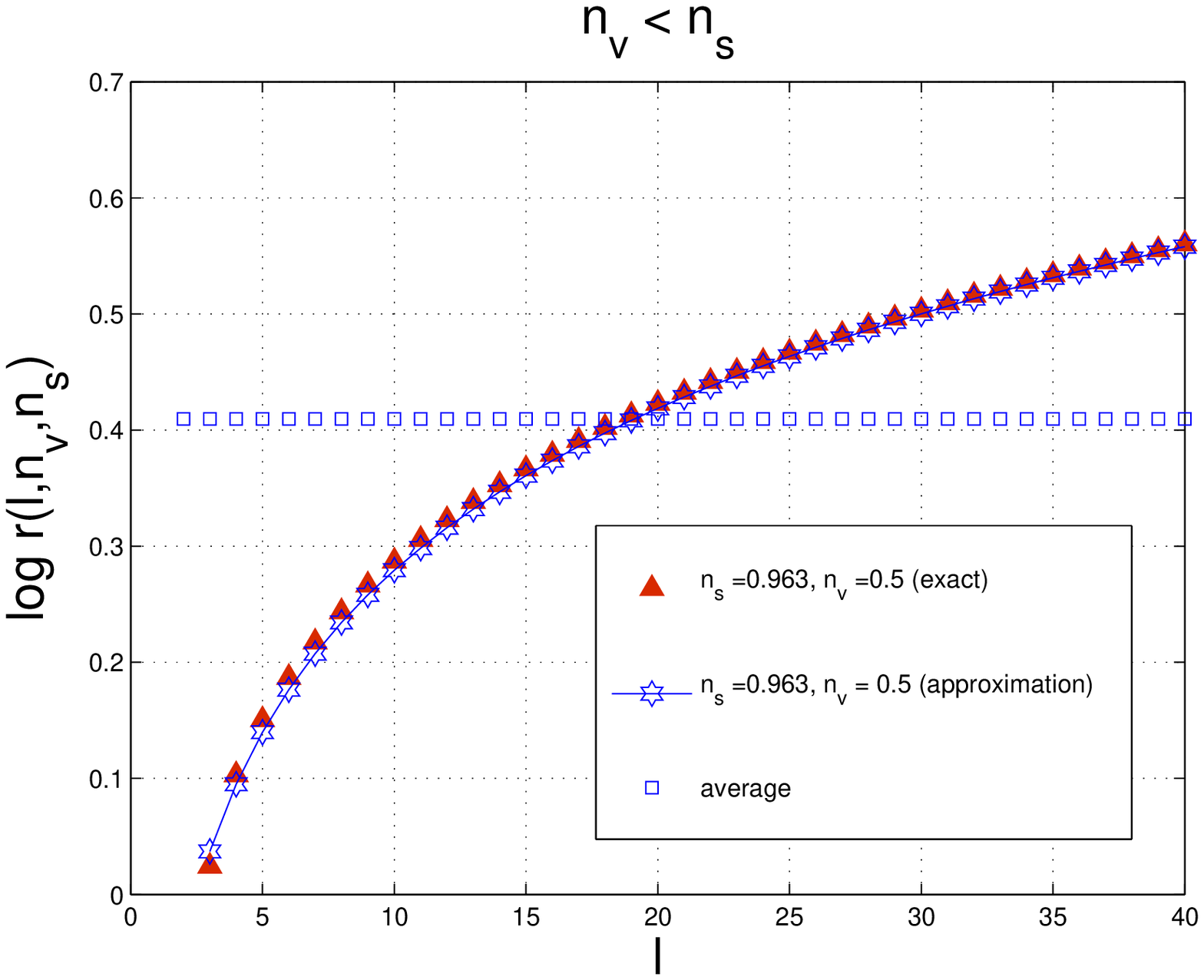}
\caption[a]{The results of Eqs. (\ref{boundTT12}) and (\ref{boundTT15}) 
are illustrated for different values of the spectral indices.}
\label{figure1}      
\end{figure}
In the range $2\leq \ell < 40$ the ratio ${\mathcal I}_{{\mathcal R}}(\ell,n_{s})/{\mathcal I}_{\mathrm{V}}(\ell,n_{v})$ 
can be usefully approximated in a factorized form as 
\begin{equation}
r(\ell, n_{v}, n_{s}) = \frac{{\mathcal I}_{{\mathcal R}}(\ell, n_{s})}{{\mathcal I}_{\mathrm{V}}(\ell, n_{s})} = \frac{4\,
\Gamma\biggl(\frac{3}{2} - \frac{n_{s}}{2}\biggr) \, \Gamma\biggl(4 - \frac{n_{v}}{2} \biggr)}{
(n_{v}^2 - 12 n_{v} + 39) \Gamma\biggl(2 - \frac{n_{s}}{2} \biggr) \Gamma\biggl(\frac{3}{2} - \frac{n_{v}}{2}\biggr)} 
\,   \ell^{n_{s} - n_{v}}\,\biggl[ 1 + {\mathcal O}\biggl(\frac{1}{\ell}\biggr)\biggr].
\label{boundTT12}
\end{equation}
The accuracy of Eq. (\ref{boundTT12}) in the physical range of spectral indices is of the order 
of few percent depending upon the value of the spectral indices and upon 
the specific multipole. The bound of Eq. (\ref{boundTT}) can now be enforced. By requiring, within our fiducial set of parameters,  that the $V$-mode contribution does not exceed the observed temperature autocorrelation we are led to the following condition
\begin{eqnarray}
{\mathcal A}_{V} < {\mathcal N}_{\mathrm{TT}} \biggl(\frac{{\mathcal A}_{{\mathcal R}}}{2.43\times 10^{-9}}\biggr) \biggl(\frac{z_{*}+ 1}{1091.79}\biggr)^{-2} 
\biggl(\frac{D_{A}}{14116\, \mathrm{Mpc}}\biggr)^{n_{v} - n_{s}} \biggl(\frac{B_{u}}{\mathrm{nG}}\biggr)^{-2} \biggl(\frac{\nu}{\mathrm{GHz}}\biggr)^{2}
\label{boundTT13}
\end{eqnarray}
where the term ${\mathcal N}_{\mathrm{TT}}$ is given by:
\begin{equation}
{\mathcal N}_{\mathrm{TT}} = 1.156 \times 10^{6} \times \,(0.0354)^{n_{s} - n_{v}} \, e^{- 2 \epsilon_{\mathrm{re}}} \, r(\ell, n_{v}, n_{s}).
\label{boundTT14}
\end{equation}
In Eq. (\ref{boundTT13}) $D_{A}$ denotes the (comoving) angular diameter distance to last scattering while $z_{*}$ denotes the redshift 
to the last scattering. All the typical values appearing in Eqs. (\ref{boundTT13}) and (\ref{boundTT14}) refer to the WMAP 7yr data alone.  It is often convenient, for sake of simplicity, to average $r(\ell, n_{v}, n_{s})$ over a suitable range of multipoles and the resulting averaged expression will be
\begin{equation}
\overline{r}(n_{v},n_{s}) = \frac{[\ell_{\mathrm{max}}^{n_{v} - n_{s} +1} -\ell_{\mathrm{min}}^{n_{v} - n_{s}+1} ]}{(\ell_{\mathrm{max}} - \ell_{\mathrm{min}})(n_{v} - n_{s} +1)} \frac{4\,
\Gamma\biggl(\frac{3}{2} - \frac{n_{s}}{2}\biggr) \, \Gamma\biggl(4 - \frac{n_{v}}{2} \biggr)}{
(n_{v}^2 - 12 n_{v} + 39) \Gamma\biggl(2 - \frac{n_{s}}{2} \biggr) \Gamma\biggl(\frac{3}{2} - \frac{n_{v}}{2}\biggr)}.
\label{boundTT15}
\end{equation}
The approximations obtained in Eqs. (\ref{boundTT12}) and (\ref{boundTT15}) 
are illustrated in Fig. \ref{figure1} where the exact results (triangles) are 
compared with the approximate result of Eq. (\ref{boundTT12}) (open stars) 
and with the average values of Eq. (\ref{boundTT15}) (open boxes) deduced from 
Eq. (\ref{boundTT15}).  The scalar spectral 
index $n_{s}$ has been taken to coincide 
with the best fit parameter of Eq. (\ref{parameters2}) while 
the spectral index $n_{v}$ has been illustrated in two specific cases, i.e. 
$n_{v} > n_{s}$ (plot at the left) and $n_{v} < n_{s}$ (plot at the right). 
The dependence upon the multipole $\ell$ can therefore 
be safely accounted for by using the derived approximate expressions and this 
trick will permit further simplifications in the final expressions.  
The current bounds on the 
V-mode polarization are rather loose for the standards we are used to in the case 
of some pivotal cosmological parameter such as the ones appearing 
in Eqs. (\ref{parameters1}), (\ref{parameters2}) and (\ref{parameters3}). 
This discussion will be approached in section \ref{sec5}. 
\renewcommand{\theequation}{4.\arabic{equation}}
\setcounter{equation}{0}
\section{E-mode autocorrelations}
\label{sec4}
In the heat transfer equation for $\Delta_{\mathrm{P}}$, there is a dependence upon 
$\Delta_{\mathrm{V}1}$ (see the right hand side of Eq. (\ref{deltaP})). 
This observation can be used to infer a further constraint on the primordial power spectrum of the circular polarization. 
For this purpose the brightness perturbation $\Delta_{\mathrm{P}}(k,\mu,\tau)$ must be appropriately related  to the E-mode polarization. In full analogy with what has been done with the TT correlations in section \ref{sec3} the idea is to compute separately the standard contribution and the V-mode contribution. 
The orthogonal combinations 
$\Delta_{\pm}(\hat{n},\tau) = \Delta_{\mathrm{Q}}(\hat{n},\tau) \pm i \Delta_{\mathrm{U}}(\hat{n},\tau)$  transform as functions of spin-weight $\pm 2$ and can therefore be expanded in terms of spin $\pm 2$ spherical harmonics \cite{sp2a,sp2b} (see 
also \cite{edm,bie,var} for a background on spin-weighted spherical harmonics):
\begin{equation}
\Delta_{\pm}(\hat{n},\tau) = \sum_{\ell \, m} a_{\pm\,2,\,\ell\, m} \, \,_{\pm\,2}Y_{\ell\, m}(\hat{n}).
\label{EM1}
\end{equation}
The E- and B-modes are, up to a sign, the real and the imaginary 
parts of $a_{\pm\,2,\ell\,m}$, i.e. 
\begin{equation}
a^{(\mathrm{E})}_{\ell\, m} = - \frac{1}{2}(a_{2,\,\ell m} + a_{-2,\,\ell m}), \qquad  
a^{(\mathrm{B})}_{\ell\, m} =  \frac{i}{2} (a_{2,\,\ell m} - a_{-2,\,\ell m}).
\label{EM2}
\end{equation}
Using the $a^{(\mathrm{E})}_{\ell\, m} $ and $a^{(\mathrm{B})}_{\ell\, m} $ it is possible to construct, in real space, $\Delta_{\mathrm{E}}(\hat{n},\tau)$ and 
$\Delta_{\mathrm{B}}(\hat{n},\tau)$ 
\begin{equation}
\Delta_{\mathrm{E}}(\hat{n},\tau) = \sum_{\ell\, m} N_{\ell}^{-1} \,  a^{(\mathrm{E})}_{\ell\, m}  \, Y_{\ell\, m}(\hat{n}),\qquad 
\Delta_{\mathrm{B}}(\hat{n},\tau) = \sum_{\ell\, m} N_{\ell}^{-1} \,  a^{(\mathrm{B})}_{\ell\, m}  \, Y_{\ell\, m}(\hat{n}),
\label{EM3}
\end{equation}
where $N_{\ell} = \sqrt{(\ell - 2)!/(\ell +2)!}$.  The 
fluctuations $\Delta_{\mathrm{E}}(\hat{n},\tau)$ and 
$\Delta_{\mathrm{B}}(\hat{n},\tau)$ are the analog of the brightness perturbations for the intensity and for the circular polarization since 
they both admit a regular expansion in terms of ordinary (i.e. spin $0$) spherical harmonics. It is practical to define a set of ladder operators raising the spin-weight of a given function. When fluctuations 
of different spin are treated on the sphere, a symmetry $O(4)$ arises naturally (see, for instance, \cite{sp2b}). Three (out of six)  generators of $O(4)$ 
correspond to the generators of the three-dimensional rotations, while the remaining three generators (commuting with the ``orbital'' angular momentum) 
are used to define the ladder operators which either raise or lower the spin-weight of a given function on the sphere. In the form which 
is suitable for the present calculation the mentioned operators read
\begin{equation}
 K_{\pm}^{\mathrm{s}}(\hat{n}) = - (\sin{\vartheta})^{\pm\mathrm{s}}\biggl[ \partial_{\vartheta} \pm  
\frac{i}{\sin{\vartheta}} \partial_{\varphi}\biggr] (\sin{\vartheta})^{\mp\mathrm{s}},
\label{EM4}
\end{equation}
By using  Eq. (\ref{EM4})  the E-mode polarization and the B-mode polarization 
are given by:
\begin{eqnarray}
&& \Delta_{\mathrm{E}}(\hat{n},\tau) = - \frac{1}{2} \{ K_{-}^{(1)}(\hat{n})[K_{-}^{(2)}(\hat{n})
\Delta_{+}(\hat{n},\tau)] +  K_{+}^{(-1)}(\hat{n})[K_{+}^{(-2)}(\hat{n}) \Delta_{-}(\hat{n},\tau)]\},
\label{EM6}\\
&&  \Delta_{\mathrm{B}}(\hat{n},\tau) =  \frac{i}{2} \{ K_{-}^{(1)}(\hat{n})[K_{-}^{(2)}(\hat{n})
\Delta_{+}(\hat{n},\tau)] -  K_{+}^{(-1)}(\hat{n})[K_{+}^{(-2)}(\hat{n}) \Delta_{-}(\hat{n},\tau)]\}.
\label{EM7}
\end{eqnarray}
In the vanilla $\Lambda$CDM scenario assumed in Eqs. (\ref{parameters1}), (\ref{parameters2}) and (\ref{parameters3}) the tensor mode power spectrum does not 
contribute; thus $\Delta_{\mathrm{B}}(\hat{n},\tau)=0$ implying, 
from Eq. (\ref{EM6}), that 
\begin{equation}
\Delta_{\mathrm{E}}(\hat{n}, \tau) = - \partial_{\mu}^2[ ( 1 - \mu^2) \Delta_{\mathrm{P}}(\hat{n}, \tau)],
\label{EM8}
\end{equation}
 where $\mu = \cos{\vartheta}$. Recalling that $\Delta_{\mathrm{P}}(\hat{n},\tau)$ obeys
(in Fourier space, Eq. (\ref{deltaP})) $a^{(\mathrm{E})}_{\ell\, m}$ can be written as 
\begin{equation}
a^{(\mathrm{E})}_{\ell\, m} = - \frac{N_{\ell}}{(2\pi)^{3/2}} \int d\, \hat{n} \, Y^{*}_{\ell\, m}(\hat{n})\, \int d^{3} k\,\partial_{\mu}^2\biggl[ ( 1 - \mu^2)\, \Delta_{\mathrm{P}}(k,\mu,\tau)\biggr].
\label{EM9}
\end{equation}
As anticipated, the solution of the equation for $\Delta_{\mathrm{P}}(k,\mu,\tau)$ contains two contributions: one stemming from $S_{\mathrm{P}}(k,\tau)$ (i.e. 
the standard adiabatic term) and one proportional to $\Delta_{\mathrm{V}1}$. 
By solving Eq. (\ref{deltaP}) with the line of sight integration we obtain:
\begin{eqnarray}
&&\Delta_{\mathrm{P}}(k,\mu,\tau_{0}) = \frac{3}{4} (1- \mu^2) \int_{0}^{\tau_{0}} d\tau 
{\mathcal K}(\tau) \, S_{\mathrm{P}}(k,\tau)\, e^{- i \mu x}
\nonumber\\
&&+ \frac{3}{2} i f_{\mathrm{e}} ( 1 - \mu^2) 
\int_{0}^{\tau_{0}} {\mathcal K}(\tau) \Delta_{\mathrm{V}1}(k,\tau),
\label{SP1}
\end{eqnarray}
where, as usual, $ x = k (\tau_{0} - \tau)$. To estimate 
 Eq. (\ref{SP1}) the  
tight coupling expansion can be used; the quadrupole of the intensity can then be related to the dipole of the intensity (but computed to zeroth order in the 
tight coupling expansion). As pointed out long ago this approximation is 
highly inaccurate  \cite{zal1}.
Instead of using the  dipole it is more accurate to obtain an equation for $S_{\mathrm{P}}(k,\tau)$, solve 
the obtained equation and then approximate the line of sight integral 
(see, e.g. \cite{zal1,mg3}). 

Since $S_{\mathrm{P}} = \Delta_{\mathrm{I}2} + \Delta_{\mathrm{P}2} + \Delta_{\mathrm{P}0}$ it is useful to derive the evolution equations of $\Delta_{\mathrm{I}2}$, $\Delta_{\mathrm{P}2}$  and $\Delta_{\mathrm{P}0}$ by taking the appropriate 
moments of the corresponding brightness perturbations, i.e. 
respectively, Eqs. (\ref{deltaI}), (\ref{deltaP}) and (\ref{deltaV}).  The result of this straightforward but lengthy manipulation is given by:
\begin{eqnarray}
 && \Delta_{{\rm P}0} ' - \frac{\epsilon'}{2} [ \Delta_{{\rm P}2} + \Delta_{{\rm I}2} - \Delta_{{\rm P}0} ]= - k \Delta_{{\rm P}1}  + i f_{\mathrm{e}} \epsilon'  \Delta_{\mathrm{V} 1},
 \label{SP2}\\
 && \Delta_{{\rm I} 2}'  + \epsilon'\biggl[ \frac{9}{10} \Delta_{{\rm I}2} - \frac{1}{10} (\Delta_{{\rm P}0} + 
\Delta_{{\rm P} 2} )\biggr] = - \frac{3}{5} k \Delta_{{\rm I}3} + \frac{2}{5} k \Delta_{{\rm I} 1} + \frac{i}{5} f_{\mathrm{e}} \epsilon'  \Delta_{\mathrm{V} 1},
\label{SP3}\\
&&  \Delta_{{\rm P} 2}' +  \epsilon'\biggl[ \frac{9}{10} \Delta_{{\rm P}2} - \frac{1}{10} (\Delta_{{\rm P}0} + 
\Delta_{{\rm I} 2} )\biggr] = - \frac{3}{5} k \Delta_{{\rm P}3} + \frac{2}{5} k \Delta_{{\rm P} 1} + i f_{\mathrm{e}} \epsilon'  \Delta_{\mathrm{V}1}.
\label{SP4}
\end{eqnarray}
Summing up Eqs. (\ref{SP2}), (\ref{SP3}) and (\ref{SP4}) the wanted equation is given by
\begin{equation}
S_{\mathrm{P}}' + \frac{3 \epsilon'}{10} S_{\mathrm{P}} = \frac{2}{5} k (\Delta_{\mathrm{I}1} + \Delta_{\mathrm{P}1}) - \frac{3}{5} k( \Delta_{\mathrm{I}3} + \Delta_{\mathrm{P}3})
+ \frac{7}{5} i \epsilon' f_{\mathrm{e}} \Delta_{\mathrm{V} 1}.
\label{SP5}
\end{equation} 
Note that, in the limit $f_{\mathrm{e}}(\omega) \to 0$ Eq. (\ref{SP5}) reproduces 
the analog equation derived in \cite{zal1}. 
The solution of Eq. (\ref{SP5}) can be obtained  with the line of sight method and the dipole of the intensity of the radiation field can be estimated to lowest order in the tight-coupling approximation.  In the standard 
case this procedure gives a rather good numerical agreement when estimating 
the polarization autocorrelations as it will be clear from the following 
considerations. The E-mode is the sum of two uncorrelated contributions 
\begin{equation}
a^{(\mathrm{E})}_{\ell\, m} = \overline{b}^{(\mathrm{E})}_{\ell\, m} + \overline{b}^{(\mathrm{V})}_{\ell\, m},
\label{EM10}
\end{equation}
whose associated power spectra are given by
\begin{eqnarray}
&& C^{(\mathrm{EE})}_{\ell} = {\mathcal B}^{(\mathrm{EE})}_{\ell} + {\mathcal B}^{(\mathrm{VV})}_{\ell},
\label{EM13}\\
&& {\mathcal B}^{(\mathrm{EE})}_{\ell} = \frac{1}{2\ell +1} \sum_{m = -\ell}^{\ell} \langle\,\, |\overline{a}^{(\mathrm{E})}_{\ell \, m}|^2\rangle,\qquad  {\mathcal B}^{(\mathrm{VV})}_{\ell} = \frac{1}{2\ell +1} \sum_{m = -\ell}^{\ell} \langle \,\, |\overline{b}^{(V)}_{\ell \, m}|^2\rangle.
\label{EM14}
\end{eqnarray}  
From the solution of Eq. (\ref{SP5}) 
in the tight-coupling approximation \cite{zal1} (see also \cite{mg3}),
the V-mode contribution to the full power spectrum is 
\begin{equation}
{\mathcal B}^{(\mathrm{VV})}_{\ell} = (4 \pi) \, f_{\mathrm{e}}^2 \, \lambda^2 \ell (\ell +1) (\ell -1) (\ell -2) 
\int \frac{d k}{k} {\mathcal P}_{V}(k) \frac{j^2_{\ell}(x)}{x^4} \cos^2{( k \, c_{\mathrm{s}v} 
\, \tau_{*})} \, e^{- 2 \frac{x^2}{\ell_{\mathrm{D}}^2}},
\label{EM15}
\end{equation}
where $\lambda = 13/2$ is a numerical factor here estimated 
to first order in the tight-coupling approximation. The diffusive damping scale $\ell_{\mathrm{D}}$ does depend upon the pivotal 
parameters of the $\Lambda$CDM scenario:
\begin{equation}
\ell_{\mathrm{D}} = k_{\mathrm{D}}\, D_{\mathrm{A}}(z_{*})= 
\frac{2240 \, d_{\mathrm{A}}(z_{*})}{\sqrt{\sqrt{r_{\mathrm{R}*} +1} - \sqrt{r_{\mathrm{R}*}}}} 
\biggl(\frac{z_{*}}{10^{3}} \biggr)^{5/4} \, \omega_{\mathrm{b}}^{0.24} \omega_{\mathrm{M}}^{-0.11},
\label{EM15a}
\end{equation}
where, following the customary notation, 
$\omega_{\mathrm{b}} = h_{0}^2\,\Omega_{\mathrm{b}0}$ and 
$\omega_{\mathrm{M}} = h_{0}^2\,\Omega_{\mathrm{M}0}$ (with $\Omega_{\mathrm{M}0} = \Omega_{\mathrm{c}0} + \Omega_{\mathrm{b}0}$). 
The (comoving) angular diameter distance at $z_{*}$ 
has been rescaled, in Eq. (\ref{EM15a}) as
\begin{equation}
D_{\mathrm{A}}(z_{*}) = \frac{2}{\sqrt{\Omega_{\mathrm{M}0}} H_{0} } d_{\mathrm{A}}(z_{*}).
\label{EM15b}
\end{equation}
Always in Eq. (\ref{EM15a}) the quantity $r_{\mathrm{R}*}$ defines ratio of the radiation and matter energy densities at $z_{*}$, i.e. 
\begin{equation}
r_{\mathrm{R}*} = \frac{\rho_{\mathrm{R}}(z_{*})}{\rho_{\mathrm{M}}(z_{*})} = \frac{a_{\mathrm{eq}}}{a_{*}}=
4.15 \times 10^{-2} \, \omega_{\mathrm{M}}^{-1}\, \biggl(\frac{z_{*}}{10^{3}}\biggr).
\label{EM15c}
\end{equation}
The numerical content of Eqs. (\ref{EM15a})--(\ref{EM15c}) is fully specified in terms of $z_{*}$ whose explicit form can be written as 
\begin{eqnarray}
z_{*} &=& 1048[ 1 + (1.24 \times 10^{-3})\, \omega_{\mathrm{b}}^{- 0.738}] [ 1 + g_{1} \omega_{\mathrm{M}}^{\,\,\,g_2}],
\label{EM15d}\\
g_{1} &=& \frac{0.0783 \, \omega_{\mathrm{b}}^{-0.238}}{[1 + 39.5 \,\,
\omega_{\mathrm{b}}^{\,\,0.763}]},\qquad 
g_{2} = \frac{0.560}{1 + 21.1 \, \omega_{\mathrm{b}}^{\,\,1.81}}.
\label{EM15e}
\end{eqnarray}
Equations (\ref{EM15d})--(\ref{EM15e}) represent a rather handy 
analytical expression for values of the parameters close to the 
$\Lambda$CDM best fit. This kind of approach has been also used 
in \cite{mg3} in a related context and  earlier  analyses can be found in 
\cite{anpar1,anpar2}. To test Eqs.  (\ref{EM15d})--(\ref{EM15e}) we can 
evaluate them  by using the fiducial set of parameters reported in 
Eqs. (\ref{parameters1})  and (\ref{parameters2}) and corresponding to the 
best fit of the WMAP 7yr data alone in the light of the vanilla $\Lambda$CDM 
paradigm (which is the one also assumed here). For instance Eqs. (\ref{EM15d}) 
and (\ref{EM15e}) imply $z_{*} =1090.77$ which is within the error bars of   
\cite{WMAP7a,WMAP7f}, i.e. $z_{*} = 1090.79^{+0.94}_{-0.92}$. Similarly 
Eq. (\ref{EM15b}) implies $D_{\mathrm{A}}(z_{*}) = 3.3397/H_{0} = 14101.6\, \mathrm{Mpc}$ again within the error bars of the WMAP 7yr data implying 
$D_{\mathrm{A}}(z_{*}) = 14116^{+160}_{-163}\, \mathrm{Mpc}$.
Equations (\ref{EM15a})--(\ref{EM15e}) depend only upon the parameters 
of the $\Lambda$CDM scenario; Eq. (\ref{EM15}) can now be rewritten by setting $ x = w \, \ell$, as
\begin{eqnarray}
{\mathcal B}^{(\mathrm{VV})}_{\ell} &=& (4 \pi) \, f_{\mathrm{e}}^2 \, \lambda^2 \ell (\ell +1) (\ell -1) (\ell -2) {\mathcal A}_{V} \biggl(\frac{k_{0}}{k_{\mathrm{p}}}\biggr)^{n_{v}-1}
\nonumber\\
&\times& \int_{1}^{\infty} \frac{d w}{w}  (\ell \, w)^{n_{v} -5} j_{\ell}^2(\ell w ) \,\cos^2{( \gamma_{v} \, \ell w)} \, e^{- 2 \frac{\ell^2}{\ell_{\mathrm{D}}^2}w^2}.
\label{EM16}
\end{eqnarray}
In the limit $\ell > 1$ and $w>1$ the spherical Bessel functions can be approximated as \cite{abr1,abr2}  
\begin{equation}
\ell (\ell +1) j^2_{\ell}( w \ell )  \simeq  \ell (\ell +1) \frac{\cos^2{[\beta(w,\ell)]}}{\ell^2 w \sqrt{w^2 - 1}} \simeq \frac{1}{2} \frac{1}{w \sqrt{w^2 -1}},  
\label{EM17}
\end{equation}
where the argument of the cosine, i.e. 
$\beta(w,\ell)= \ell \sqrt{w^2 -1} - \ell\arccos{(1/w )} - \frac{\pi}{4}$, leads to rapidly 
oscillating contributions averaging to $1/2$ in the final expression. Similar 
techniques are employed in the semi-analytical discussion of the temperature 
autocorrelations \cite{mg3} (see also \cite{wein,mg3a}).
 After some algebra ${\mathcal B}^{(\mathrm{VV})}_{\ell}$ becomes  
\begin{eqnarray}
&& \frac{\ell (\ell +1)}{2\pi} {\mathcal B}^{(\mathrm{VV})}_{\ell}
 = \lambda^2 f_{\mathrm{e}}^2 \, 
{\mathcal A}_{V} (\ell + \ell_{V})^{n_{v} -1} \biggl(\frac{k_{0}}{k_{\mathrm{p}}}\biggr)^{n_{v} -1} [ {\mathcal P}_{\ell}(\ell_{\mathrm{D}}, n_{v}) + {\mathcal Q}_{\ell}(\ell_{\mathrm{D}}, n_{v})],
\label{EM18}
\end{eqnarray}
where 
\begin{eqnarray}
&& {\mathcal P}_{\ell}(\ell_{\mathrm{D}}, n_{v}) =  \int_{1}^{\infty} d w \frac{w^{n_{v} -7}}{\sqrt{w^2 -1}} e^{- 2 \frac{\ell^2}{\ell_{\mathrm{D}}^2} w^2 },
\label{EM18a}\\
&& {\mathcal Q}_{\ell}(\ell_{\mathrm{D}}, n_{v}) =  \int_{1}^{\infty} d w \frac{w^{n_{v} -7}}{\sqrt{w^2 -1}} \cos{( 2 \gamma_{v} \ell w)} e^{- 2 \frac{\ell^2}{\ell_{\mathrm{D}}^2} w^2 }.
\label{EM18b}
\end{eqnarray}
The integrals appearing in Eqs. (\ref{EM18a}) and (\ref{EM18b}) shall be performed 
numerically by changing, for instance, the integration variable as $w = \sqrt{y^2 +1}$: 
\begin{eqnarray}
{\mathcal P}_{\ell}(\ell_{\mathrm{D}}, n_{v}) = \frac{1}{2}\, e^{- 2(\ell^2/\ell_{\mathrm{D}})^2} \int_{0}^{\infty} (y^2 +1)^{(n_{v}-8)/2}  \, e^{ - 2 (\ell^2/\ell_{\mathrm{D}})^2\, y^2},
\label{EM18c}\\
{\mathcal Q}_{\ell}(\ell_{\mathrm{D}}, n_{v}) = \frac{1}{2}\, e^{- 2(\ell^2/\ell_{\mathrm{D}})^2} \int_{0}^{\infty} (y^2 +1)^{(n_{v}-8)/2} \,\cos{(2\ell \gamma_{v} \sqrt{y^2 +1})} \, e^{ - 2 (\ell^2/\ell_{\mathrm{D}})^2\, y^2}.
\label{EM18d}
\end{eqnarray}
Interestingly enough, the final result can be parametrized as
\begin{eqnarray}
&&\frac{\ell (\ell +1)}{2\pi} {\mathcal B}^{(\mathrm{VV})}_{\ell} = {\mathcal N}_{\mathrm{VV}} (\ell + \ell_{\mathrm{V}})^{n_{v} -1}\biggl\{ a_{\mathrm{V}} + b_{\mathrm{V}} 
\cos{[2 \gamma_{v} (\ell + \ell_{\mathrm{V}})]} \biggr\} e^{-2 (\ell/\ell_{\mathrm{D}})^2}, 
\label{EM18e}\\
&& {\mathcal N}_{\mathrm{VV}}= \lambda^2 f_{\mathrm{e}}^2(\omega)  \, 
{\mathcal A}_{V}  \biggl(\frac{k_{0}}{k_{\mathrm{p}}}\biggr)^{n_{v} -1} \, T_{\gamma 0}^2
\label{EM18f}\\
&& a_{\mathrm{V}} =0.54,\qquad b_{\mathrm{V}} = 0.22,\qquad \ell_{\mathrm{V}} = 65,
\label{EM18g}
\end{eqnarray}
where $\gamma_{v} = (\tau_{*}/\tau_{0}) c_{\mathrm{s}v}\simeq 0.06/\sqrt{3}$.
The obtained results should be compared 
with the corresponding expression for the ${\mathcal B}_{\ell}^{(\mathrm{EE})}$ power spectrum. Modulo the difference in the fiducial set of parameters (which 
coincide here with the ones of Eqs. (\ref{parameters1}) and (\ref{parameters2}))
the result can be read off directly from \cite{mg3} 
\begin{eqnarray}
&&\frac{\ell (\ell +1)}{2\pi} {\mathcal B}_{\ell}^{(\mathrm{EE})}=
{\mathcal N}_{(\mathrm{EE})} (\ell + \ell_{\mathrm{E}})^{n_{s}+1}\,\biggl\{ a_{\mathrm{E}} - b_{\mathrm{E}} \cos{[2 \gamma_{\mathrm{A}} (\ell + \ell_{\mathrm{E}})]}\biggr\}\, e^{- 2 (\ell/\ell_{\mathrm{D}})^2},
\label{EE22a}\\
&& {\mathcal N}_{\mathrm{EE}} = 4.453 \times 10^{-4} \, \biggl(\frac{k_{0}}{k_{\mathrm{p}}}\biggr)^{n_{s} -1} \, \biggl(\frac{{\mathcal A}_{{\mathcal R}}}{2.43\times 10^{-9}}\biggr) e^{- 2 \epsilon_{\mathrm{re}}} \, (\mu\mathrm{K})^2,
\label{EE22}\\
&& a_{\mathrm{E}}= 0.194,\qquad b_{\mathrm{E}}=0.110, \qquad \ell_{\mathrm{E}} = 65.
\label{EE22b}
\end{eqnarray}
The fact that $\ell_{\mathrm{E}} \simeq \ell_{\mathrm{V}}$ is a consequence 
of the use of the same asymptotic expression of the spherical Bessel functions 
in both sets of integrals. In Eq. (\ref{EE22a}) $\gamma_{\mathrm{A}}$ is nothing but $\pi/\ell_{\mathrm{A}}$ where $\ell_{\mathrm{A}}$ is the acoustic monopole, i.e. 
\begin{equation} 
\ell_{\mathrm{A}} = \biggl(\frac{z_{*}}{10^{3}}\biggr)^{1/2} \frac{\sqrt{R_{\mathrm{b}*}}\,d_{\mathrm{A}}(z_{*})}{\ln{\biggl[
\frac{\sqrt{1 + R_{\mathrm{b}*}} + \sqrt{( 1 +  r_{\mathrm{R}*})R_{\mathrm{b}*}}}{1 + 
\sqrt{r_{\mathrm{R}*} R_{\mathrm{b}*}}}\biggr]}},
\label{EE22c}
\end{equation} 
where $R_{\mathrm{b}}(z_{*})$ is the baryon-photon ratio, $c_{\mathrm{sb}}(z_{*})$ 
is baryon-photon sound speed and 
\begin{eqnarray}
&& c_{\mathrm{sb}}(z_{*}) = \frac{1}{\sqrt{3[ 1 + R_{\mathrm{b}}(z_{*})]}}, \qquad  
R_{\mathrm{b}}(z_{*}) = \frac{3}{4} \frac{\rho_{\mathrm{b}}}{\rho_{\gamma}} = 
30.36 \,\omega_{\mathrm{b}} \,\biggl(\frac{10^{3}}{z_{*}}\biggr),
\label{EE22d}\\
&& r_{\mathrm{s}}(z_{*}) = \int_{0}^{\tau_{*}} d\tau \, c_{\mathrm{s}\mathrm{b}}(\tau) = 
 \int_{0}^{\tau_{*}} \,\frac{d\tau}{\sqrt{3 [ R_{\mathrm{b}}(\tau) +1]}}.
\label{EE22e}
\end{eqnarray}
The parametrization of Eq. (\ref{EE22c}) implies that 
$\ell_{\mathrm{A}} = 301.674$ which is consistent with 
 the WMAP 7yr data giving $\ell_{\mathrm{A}} = 302.57^{+0.77}_{-0.76}$.
\begin{figure}[!ht]
\centering
\includegraphics[height=6cm]{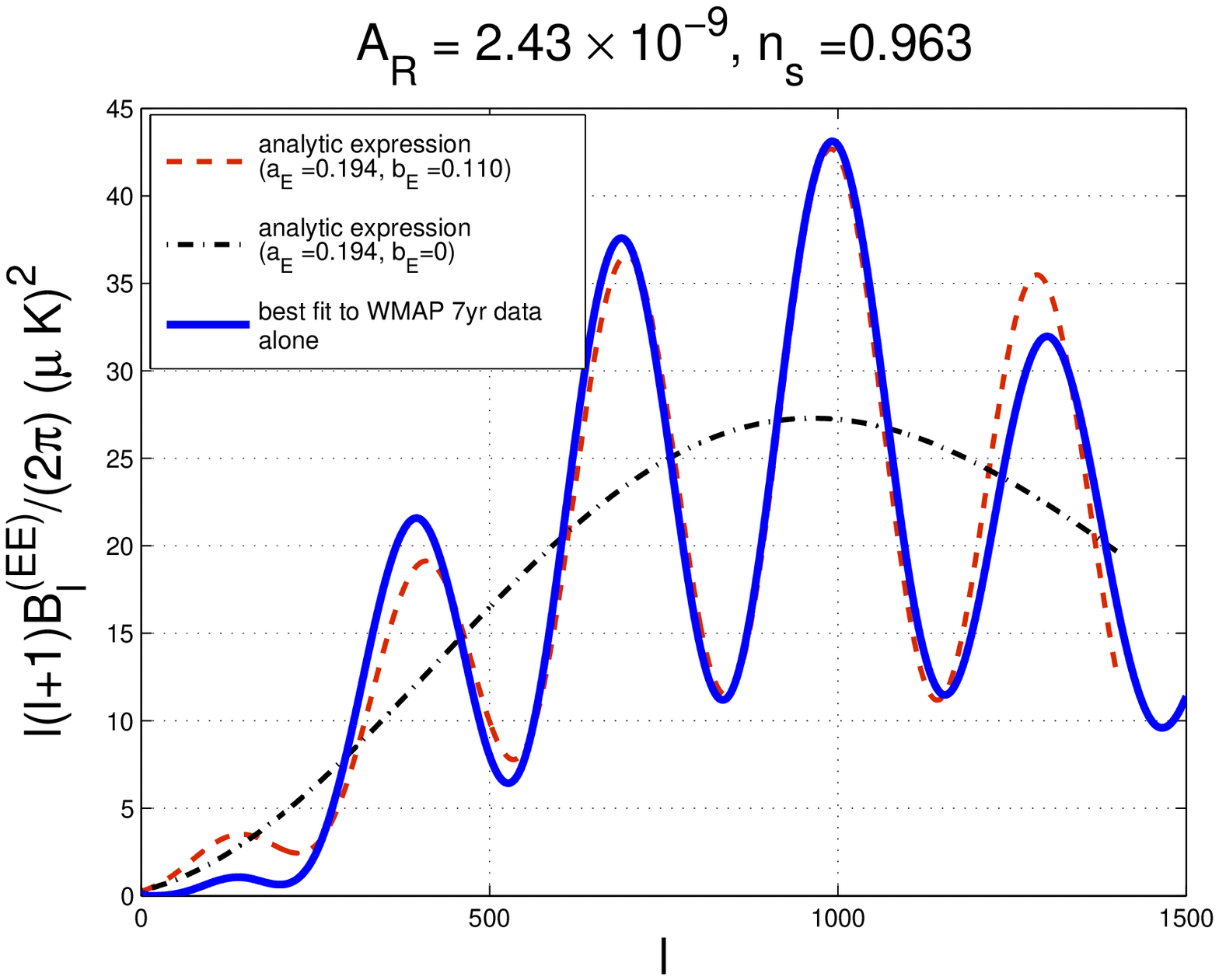}
\includegraphics[height=6cm]{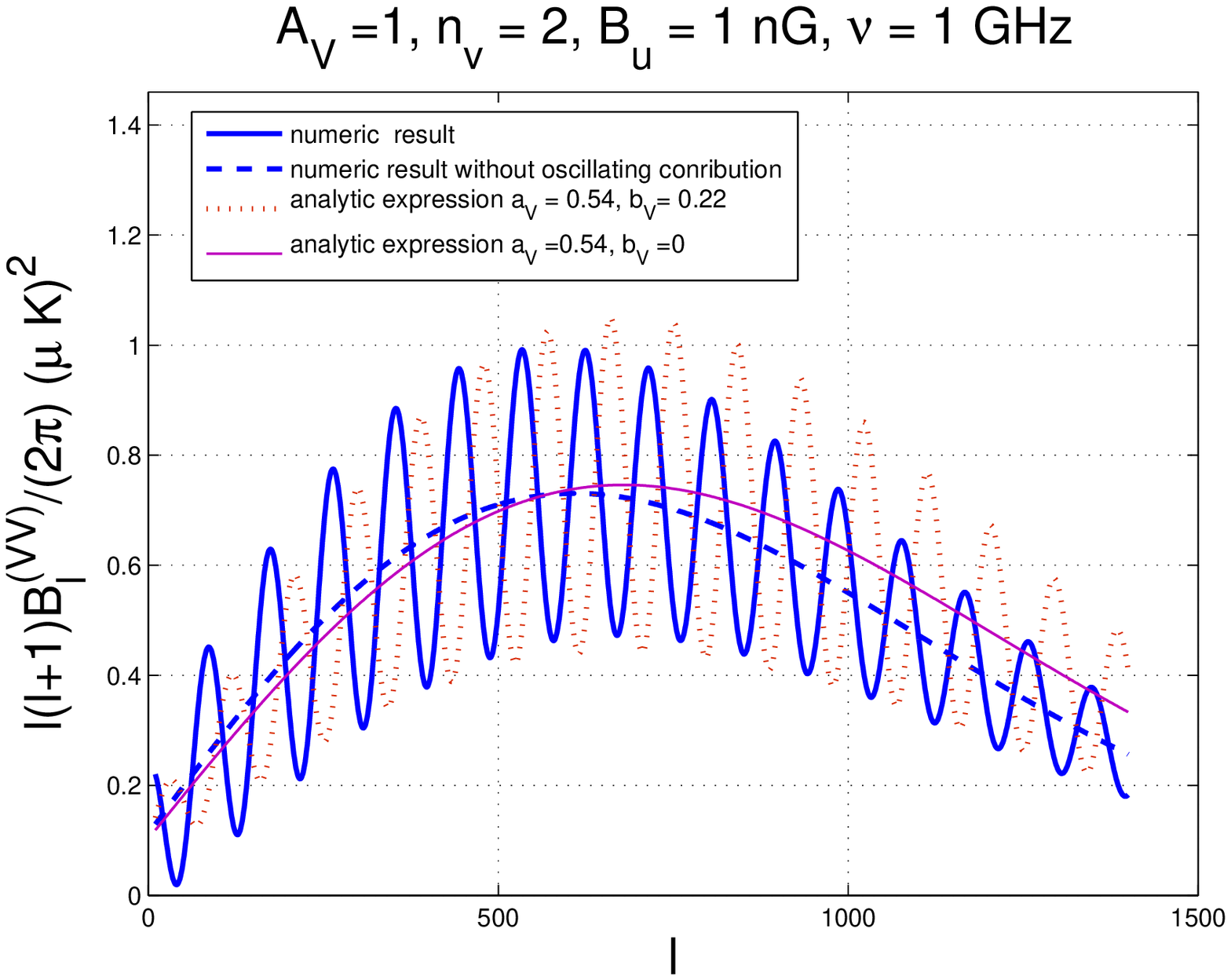}
\caption[a]{The semi-analytical results of Eqs. (\ref{EE22a})--(\ref{EE22b}) and of 
Eqs. (\ref{EM18e})--(\ref{EM18g}) are illustrated, respectively, in the 
left and in the right panels. Since ${\mathcal A}_{\mathrm{V}}$ is the quantity 
we ought to bound, in the plot at the right arbitrary units 
${\mathcal A}_{\mathrm{V}} =1$ have been employed just for illustration.}
\label{figure2}      
\end{figure}
In Fig. \ref{figure2} the various approximations discussed so far in this section 
are summarized. 

In the plot at the left of Fig. \ref{figure2} the full line illustrates 
the EE correlation computed numerically using the best fit parameters 
derived from the WMAP 7yr data (see Eqs. (\ref{parameters1}) and (\ref{parameters2})).
In the same plot the dashed line denotes the semi-analytical result 
of Eq. (\ref{EE22a}). Finally, always in the plot at the left, the dot-dashed 
line is the graphical illustration of the result of Eq. (\ref{EE22a}) but with 
$b_{\mathrm{E}} =0$. The maximum of the dot-dashed curve is located for 
\begin{equation}
\ell_{\max}^{(\mathrm{E})} = \frac{\sqrt{\ell_{\mathrm{E}}^2 + (n_{s} +1) \ell_{\mathrm{D}}^2} - \ell_{\mathrm{E}}}{2}.
\label{EE23}
\end{equation}
By evaluating Eq. (\ref{EE22a}) in for $\ell= \ell_{\max}^{(\mathrm{E})}$ 
the maximum of $\ell (\ell +1) {\mathcal B}^{(\mathrm{EE})}/(2\pi)$ 
can be estimated. In the plot at the right, always in Fig. \ref{figure2},
the V-mode contribution is illustrated with the same logic used in 
the plot at the left: the semi-analytical expression of Eq. (\ref{EM18a}) 
is compared with the numerical results. The maximum of 
$\ell (\ell +1) {\mathcal B}^{(\mathrm{VV})}/(2\pi)$ is located for 
\begin{equation}
\ell_{\max}^{(\mathrm{V})} = \frac{\sqrt{\ell_{\mathrm{V}}^2 + (n_{v} -1) \ell_{\mathrm{D}}^2} - \ell_{\mathrm{V}}}{2}.
\label{EE24}
\end{equation}
To obtain the coveted analytical bound on the V-mode contribution it is thus
then sufficient to require that: 
\begin{equation}
\biggl[ \frac{\ell (\ell +1) {\mathcal B}^{(\mathrm{VV})}}{2\pi} \biggr]_{\ell = \ell_{\max}^{(\mathrm{V})}} < \biggl[ \frac{\ell (\ell +1) {\mathcal B}^{(\mathrm{EE})}}{2\pi} \biggr]_{\ell = \ell_{\max}^{(\mathrm{E})}}.
\label{EE25}
\end{equation}
By enforcing the condition given by Eq. (\ref{EE25}) we obtain:
\begin{eqnarray}
&& {\mathcal A}_{\mathrm{V}} < 0.151\times \frac{a_{\mathrm{E}}}{a_{\mathrm{V}}}\, \sqrt{\frac{(n_{s} +1)^{n_{s} +1}}{(n_{v} -1)^{n_{v} -1}}} \, \biggl( \frac{4 e}{\ell_{\mathrm{D}}}\biggr)^{(n_{v} - n_{s})/2 -1}
\nonumber\\
&& \times \biggl(\frac{{\mathcal A}_{{\mathcal R}}}{2.43\times 10^{-9}}\biggr) \biggl(\frac{z_{*}+ 1}{1091.79}\biggr)^{-2} 
\biggl(\frac{D_{A}(z_{*})}{14116\, \mathrm{Mpc}}\biggr)^{n_{v} - n_{s}} \biggl(\frac{B_{u}}{\mathrm{nG}}\biggr)^{-2} \biggl(\frac{\nu}{\mathrm{GHz}}\biggr)^{2},
\label{EE26}
\end{eqnarray}
for $n_{v} >1$ and
\begin{eqnarray}
&& {\mathcal A}_{\mathrm{V}} < 0.151\times \frac{a_{\mathrm{E}}}{a_{\mathrm{V}}} \, (2 \ell_{\mathrm{V}})^{1 - n_{v}}\, 
(n_{s} +1)^{(n_{s} +1)/2} \, \biggl( \frac{4 e}{\ell_{\mathrm{D}}}\biggr)^{-( n_{s} +1)/2}
\nonumber\\
&& \times \biggl(\frac{{\mathcal A}_{{\mathcal R}}}{2.43\times 10^{-9}}\biggr) \biggl(\frac{z_{*}+ 1}{1091.79}\biggr)^{-2} 
\biggl(\frac{D_{A}(z_{*})}{14116\, \mathrm{Mpc}}\biggr)^{n_{v} - n_{s}} \biggl(\frac{B_{u}}{\mathrm{nG}}\biggr)^{-2} \biggl(\frac{\nu}{\mathrm{GHz}}\biggr)^{2},
\label{EE27}
\end{eqnarray}
for $n_{v} < 1$. 
\renewcommand{\theequation}{5.\arabic{equation}}
\setcounter{equation}{0}
\section{Limits on the V-mode autocorrelations}
\label{sec5}
The contribution of the circular polarization to the TT and to the EE correlations has been separately computed in the two previous sections. Two different sets of bounds 
have been derived from complementary considerations. It is now interesting to summarize the whole discussion in the light possible (direct) experimental limits on the V-mode autocorrelation.
Current bounds on circular polarization coming from direct searches are, to the best of our knowledge, still the ones 
given in \cite{lubin1} (see also \cite{lubin2,lubin3}) and in \cite{part1} (see also 
\cite{part2,part3}). The measurements of  \cite{part1,part2,part3}
 were conducted for a typical wavelength of $6$ cm (corresponding to $\nu= 4.9$ GHz) and used the Very Large Array radio-telescope in Socorro (New Mexico).
 Conversely the limits of \cite{lubin1,lubin2,lubin3} used a $\nu=33$ GHz radiometer
 (corresponding to a wavelength of $9$ mm)  which used a Faraday rotator to switch 
 between orthogonal and linear polarization states. 
 Besides the difference in frequency the two experiments also probed different angular 
 scales. The experiment of Ref. \cite{lubin1}  was sensitive to pretty large 
 angular scales (i.e. $\vartheta \simeq {\mathcal O}(10)$ deg) and obtained 
\begin{equation}
\sqrt{\frac{\ell (\ell +1)}{2\pi} C_{\ell}^{(\mathrm{VV})}} \leq \alpha_{1}\, \mathrm{m K},
\label{limit1}
\end{equation}
where $1\, \mathrm{mK} = 10^{-3}\, \mathrm{K}$ and 
 $\alpha_{1}$ ranges from $20$ to $0.2$ (if we consider 
the sensitivity per each beam patch of $7$ deg).
The results of \cite{part1,part2,part3} hold instead for a much smaller range of angular scales ranging from $18$ arc s to $160$ arc s. The authors of \cite{part1} report their results for the amplitude in terms of the equivalent temperature 
fluctuation (for which only upper limits existed at that time) and in terms of a 
putative CMB temperature of $2.75$ K. In terms of these quantities 
we can establish, within the notation of Eq. (\ref{limit1}) that 
\begin{equation}
\sqrt{\frac{\ell (\ell +1)}{2\pi} C_{\ell}^{(\mathrm{VV})}} \leq \alpha_{2}\, \mathrm{m K},
\label{limit2}
\end{equation}
where, now, $\alpha_{2}$ is given by 
\begin{eqnarray}
&& \alpha_{2} = 0.605 ,\qquad 18 \, \mathrm{arc\,s} < \vartheta < 160 \, \mathrm{arc\,s},
\label{limit3}\\
&& \alpha_{2} = 0.286 ,\qquad 36 \, \mathrm{arc\,s} < \vartheta < 160 \, \mathrm{arc\,s},
\label{limit4}\\
&&\alpha_{2} = 0.173 ,\qquad 60 \, \mathrm{arc\,s} < \vartheta < 160 \, \mathrm{arc\,s}.
\label{limit5}
\end{eqnarray}
The conversion between different notations can be  performed 
by noticing that the two-point function for the V-mode in real space is given by
\begin{equation}
\langle \Delta_{\mathrm{V}}(\hat{n}_{1})  \Delta_{\mathrm{V}}(\hat{n}_{2}) \rangle = 
\sum_{\ell} \frac{(2\ell +1)}{4\pi} C_{\ell}^{(\mathrm{VV})} P_{\ell}(\vartheta),\qquad \vartheta = \hat{n}_{1} \cdot \hat{n}_{2}
\label{limit6}
\end{equation}
and by recalling that $\ell(\ell+1)C_{\ell}^{(\mathrm{VV})}/(2\pi)$, measures, 
effectively, the amplitude of the two-point function per logarithmic interval 
of multipole:
\begin{equation}
\sum_{\ell} \frac{2\ell +1}{4\pi} C_{\ell}^{(\mathrm{VV})}\simeq \int \frac{d \ell}{\ell} \, 
\frac{\ell (\ell +1)}{2\pi} \, C_{\ell}^{(\mathrm{VV})}.
\label{limit7}
\end{equation}
The limits listed in Eq. (\ref{limit1}) and in  Eqs. (\ref{limit3}), (\ref{limit4}) and 
(\ref{limit5}) are not so recent and we are therefore confronted with various potential 
ambiguities related not only to the angular scale and to the frequency channel 
but also to the very values of the CMB temperature which differ from the ones 
determined in the framework of the WMAP 7 data release. To cope with this 
situation the following parametrization will be adopted for the direct limits on the 
V-mode power spectrum:
\begin{equation}
\sqrt{\frac{\ell (\ell +1)}{2\pi} C_{\ell}^{(\mathrm{VV})}} = \alpha \, T_{\gamma 0}, \qquad T_{\gamma 0} =2.725 \, \mathrm{K}.
\label{limit8}
\end{equation}
Different values of $\alpha$ will correspond to different observational limits either 
already obtained or potentially interesting for the present considerations. It is actually scarcely disputable that the observational limits on the V-mode 
polarization quoted in \cite{lubin1,lubin2,lubin3} and in \cite{part1,part2,part3} 
can and should be improved. 
Using the line of sight integration, Eq. (\ref{deltaV}) implies
\begin{eqnarray}
&& a_{\ell \, m}^{(\mathrm{V})} = \frac{1}{(2\pi)^{3/2}} \int \, d\hat{n}\, Y_{\ell\, m}^{*}(\hat{n}) \, \int d^{3} k\, \Delta_{\mathrm{V}}(k,\mu,\tau_{0}),
\label{VM1}\\
&& \Delta_{V}(k,\mu,\tau_{0}) =  -\frac{3}{4}\,i \, \mu \int_{0}^{\tau_{0}} d\tau \, {\mathcal K}(\tau) \, e^{- i \mu x} \Delta_{\mathrm{V}\,1}(k,\tau). 
\label{VM2}
\end{eqnarray}
The limits on small angular scales on the V-mode polarization are of the same 
order of the limits obtained for larger angular scales. 
The V-mode autocorrelation can be neatly computed in the sudden decoupling limit 
where  the coefficient $a_{\ell\, m}^{(\mathrm{V})}$ can be written as 
\begin{equation}
a_{\ell\, m}^{(\mathrm{V})} = \frac{3\, (-i)^{\ell}}{4 \, (2 \pi)^{3/2}} \delta_{m0} \sqrt{\frac{4\pi}{2 \ell + 1}} \int d^{3} k \, \int_{0}^{\tau_{0}}
 [ \ell \,j_{\ell -1}(x) - (\ell + 1) \, j_{\ell +1}(x) ]\, \Delta_{\mathrm{V}\,1}(k,\tau) \, d\tau;
 \label{VM3}
 \end{equation}
  as usual, $ x = k (\tau_{0} - \tau)$. In the large-scale limit (i.e., in practice for $\ell < 40$) the angular power spectrum 
 of the V-mode polarization is given by:
 \begin{equation}
 C_{\ell}^{(\mathrm{VV})} =  \frac{9\pi}{4 \, ( 2 \ell+ 1)^2} \int_{0}^{\infty} \frac{d k}{k} \, {\mathcal P}_{V}(k) [\ell j_{\ell-1}(x) - (\ell +1) j_{\ell +1}(x) ]^2. 
 \label{VM4}
 \end{equation}
 \begin{figure}[!ht]
\centering
\includegraphics[height=6cm]{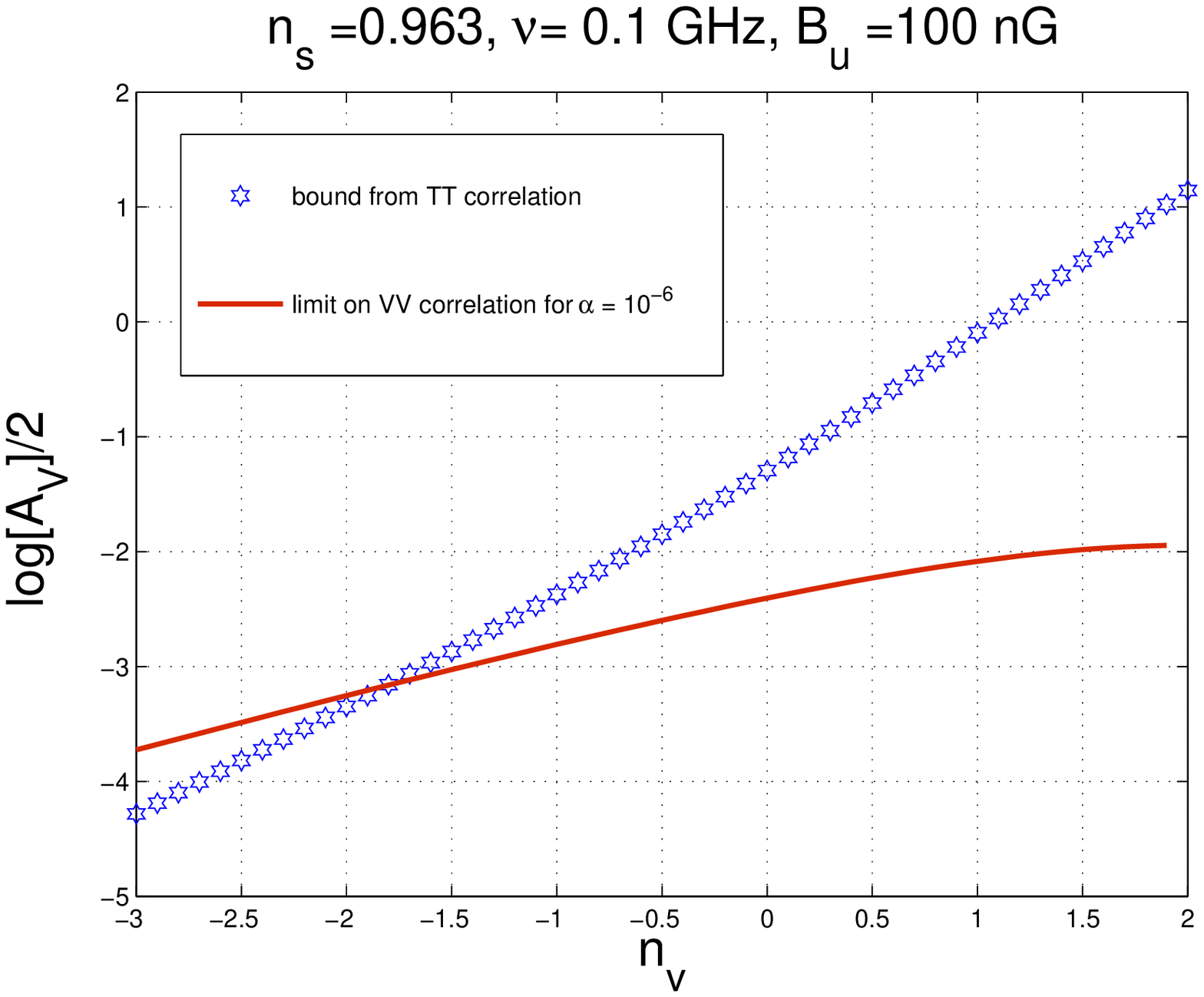}
\includegraphics[height=6cm]{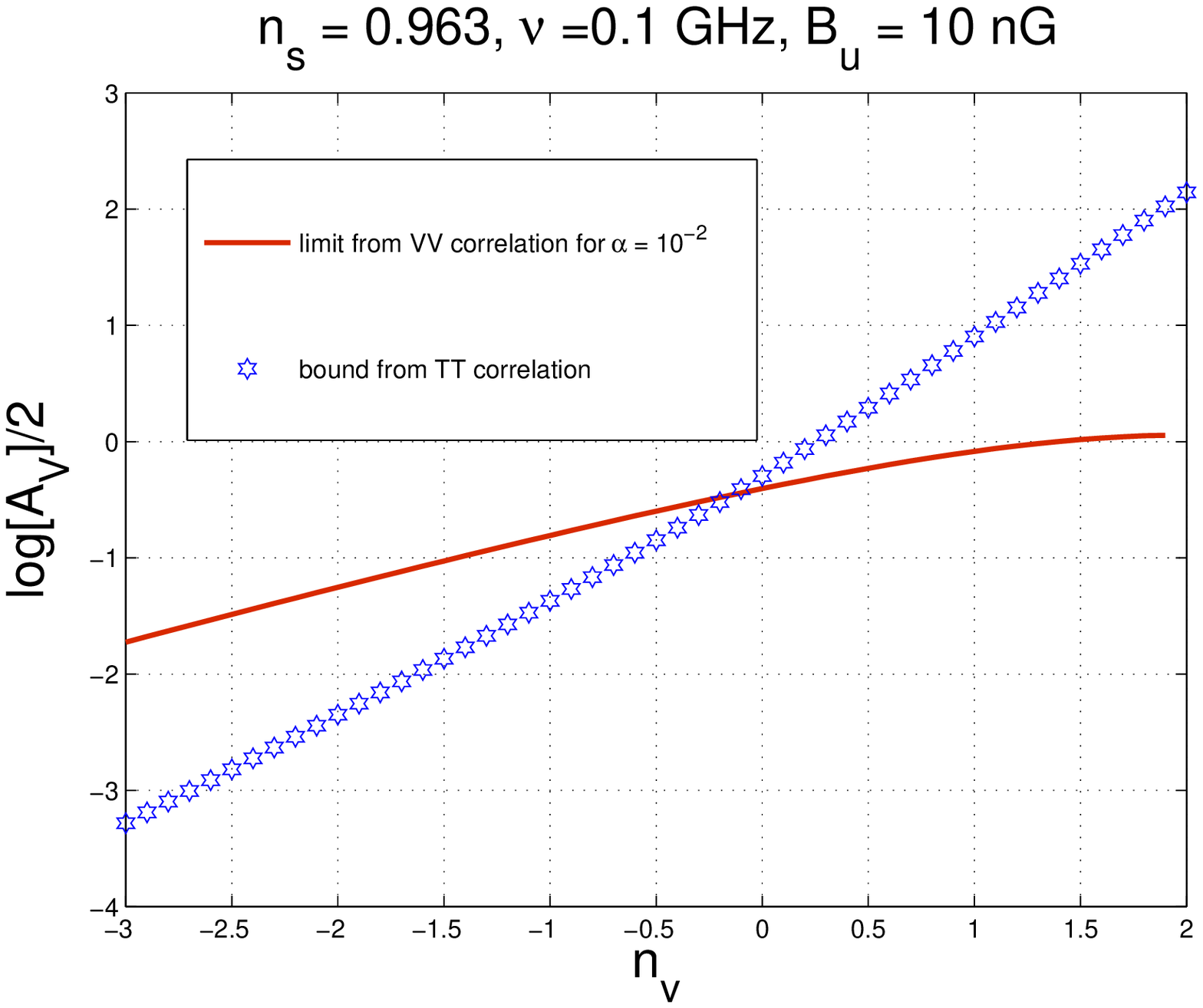}
\includegraphics[height=6cm]{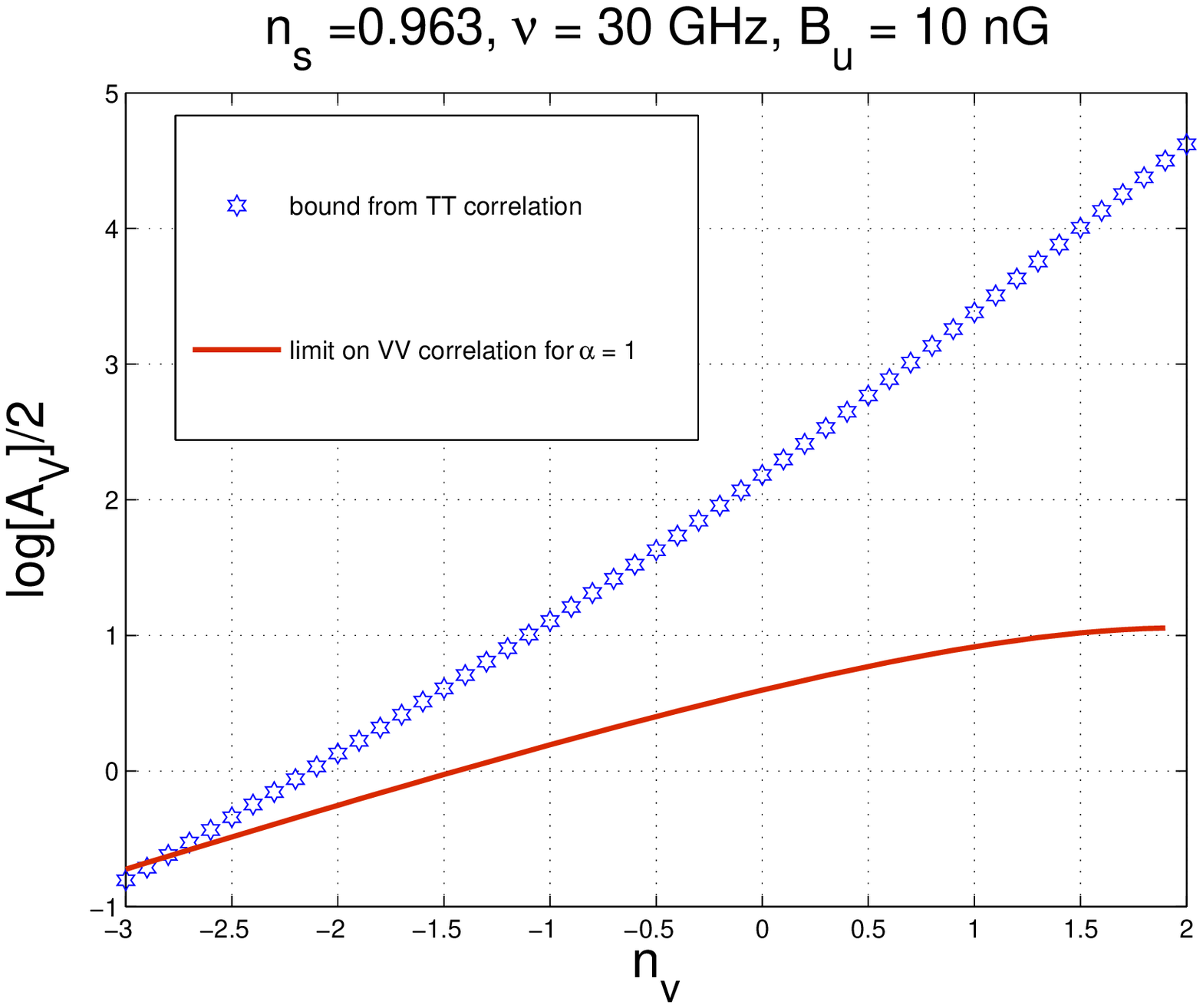}
\includegraphics[height=6cm]{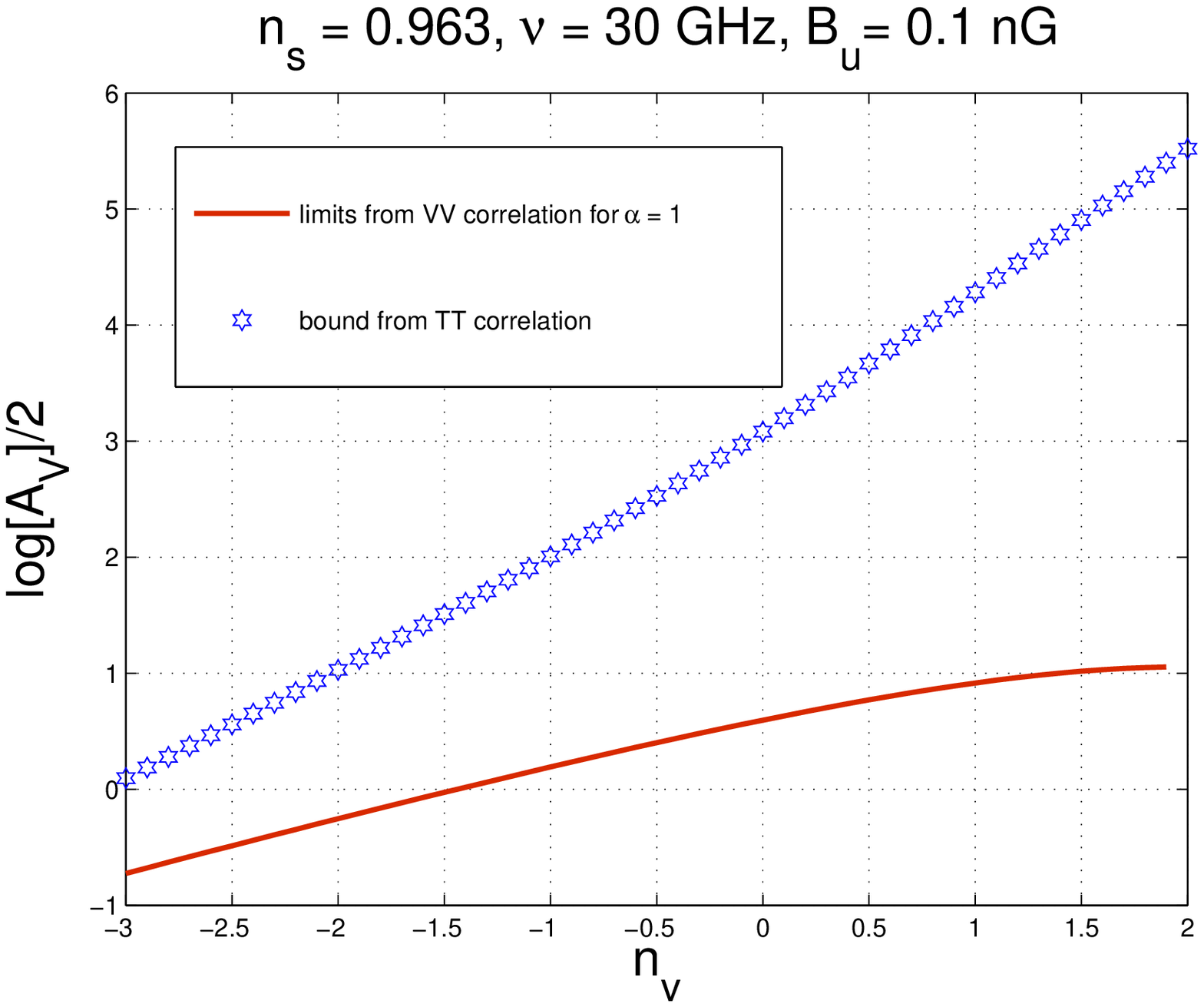}
\caption[a]{The bounds on the V-mode power spectrum are illustrated 
in tems of the amplitude and of the spectral index. The starred points correspond 
to the bounds arising from  Eqs. (\ref{boundTT13}) and (\ref{boundTT14}).
The full lines corresponds to Eq. (\ref{limit8}) with the values of $\alpha$ 
reported in each legend.}
\label{figure4}      
\end{figure}
The latter expression can be explicitly computed and the result is 
\begin{eqnarray}
&& C_{\ell}^{(\mathrm{VV})} = \frac{9 \pi^2}{8} {\mathcal A}_{\mathrm{V}} \, \biggl(\frac{k_{0}}{k_{\mathrm{p}}}\biggr)^{n_{v} -1} {\mathcal V}(\ell, n_{v}),
\label{VM5}\\
&& {\mathcal V}(\ell,n_{v}) = \frac{\ell^2}{(2 \ell +1)^2} {\mathcal V}_{1}(\ell, n_{v}) + \frac{(\ell +1)^2}{(2\ell +1)^2} {\mathcal V}_{2}(\ell, n_{v}) 
- \frac{2 \ell (\ell +1)}{(2\ell +1)^2} {\mathcal V}_{3}(\ell, n_{v}),
\label{VM6}
\end{eqnarray}
where the functions ${\mathcal V}_{1}(\ell, n_{v})$, ${\mathcal V}_{2}(\ell, n_{v})$ and ${\mathcal V}_{3}(\ell, n_{v})$ are given by 
\begin{eqnarray}
&&{\mathcal V}_{1}(\ell,n_{v}) = \int_{0}^{\infty} dx \, x^{n_{v}-3} \, J^2_{\ell - 1/2}(x) = \frac{1}{2 \sqrt{\pi}} \frac{\Gamma\biggl(\frac{3}{2} - \frac{n_{v}}{2}\biggr)
\Gamma\biggl(\ell - \frac{3}{2} + \frac{n_{v}}{2} \biggr)}{\Gamma\biggl(2 - \frac{n_{v}}{2}\biggr)\, \Gamma\biggl(\frac{3}{2} + \ell - \frac{n_{v}}{2} \biggr)},
\label{VM7}\\
&& {\mathcal V}_{2}(\ell,n_{v}) = \int_{0}^{\infty} dx \, x^{n_{v}-3} \, J^2_{\ell + 3/2}(x) = \frac{1}{2 \sqrt{\pi}} \frac{\Gamma\biggl(\frac{3}{2} - \frac{n_{v}}{2}\biggr) \Gamma\biggl(\ell - \frac{1}{2} + \frac{n_{v}}{2}\biggr)}{\Gamma\biggl(2 - \frac{n_{v}}{2} \biggr) \Gamma\biggl(\frac{7}{2} +\ell - \frac{n_{v}}{2} \biggr)},
\label{VM8}\\
&& {\mathcal V}_{3}(\ell,n_{v}) = \int_{0}^{\infty} dx \, x^{n_{v}-3} \, J_{\ell + 3/2}(x)  J_{\ell -1/2}(x) = \frac{(2- n_{v})}{4 \sqrt{\pi}} \frac{\Gamma\biggl(\frac{3}{2} - \frac{n_{v}}{2}\biggr) \Gamma\biggl(\ell - \frac{1}{2} + \frac{n_{v}}{2} \biggr)}{\Gamma\biggl(3 - \frac{n_{v}}{2} \biggr) \Gamma\biggl(\frac{5}{2} + \ell - \frac{n_{v}}{2} \biggl)}.
\label{VM9}
\end{eqnarray}
As previously done,  it is practical to deduce a simplified expression valid in the limit $\ell > 1$:
\begin{equation}
{\mathcal V}(\ell,n_{v}) = \frac{\ell^{n_{v} -3}}{2 \sqrt{\pi} ( 4 - n_{v})} \frac{\Gamma\biggl(\frac{3 - n_{v}}{2}\biggr)}{\Gamma\biggl(\frac{4 - n_{v}}{2}\biggr)}\biggl[ 1 + {\mathcal O}\biggl(\frac{1}{\ell}\biggr)\biggr].
\label{VM10}
\end{equation}
Since ${\mathcal A}_{\mathrm{V}}$ has been independently bounded from the analysis of the TT and of the EE angular power spectra, the $V$-mode angular power 
spectrum is also bounded. 
In Fig. \ref{figure4} the bounds stemming from the V-mode contribution to the TT 
power spectrum are summarized for different values 
of the magnetic field intensity. In all four plots on the vertical axis we report 
the common logarithm of $\sqrt{{\mathcal A}_{\mathrm{V}}}$ 
while on the horizontal axis the corresponding spectral index is illustrated. 

To derive the plots of Fig. \ref{figure4} it has been assumed that all the parameters 
of the (vanilla) $\Lambda$ CDM scenario are fixed to the values 
of Eqs. (\ref{parameters1})--(\ref{parameters3}). 
The latter statement simply stipulates that 
the V-mode polarization does not affect directly the determinations of the 
$\Lambda$CDM parameters at least in the first approximation. The frequency 
channel is part of the observational set-up and it is therefore fixed by the 
nature of the experimental apparatus. 
\begin{figure}[!ht]
\centering
\includegraphics[height=6cm]{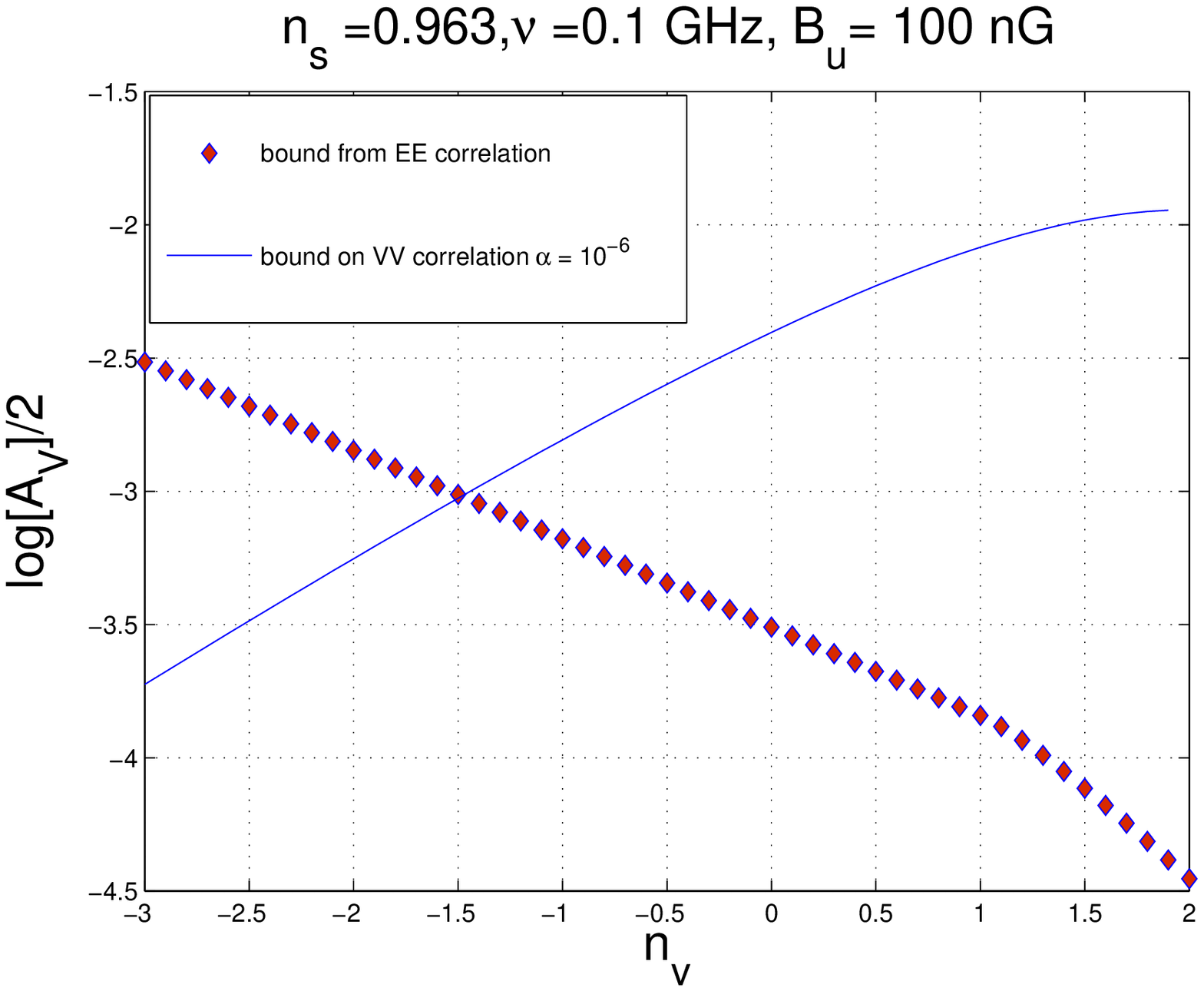}
\includegraphics[height=6cm]{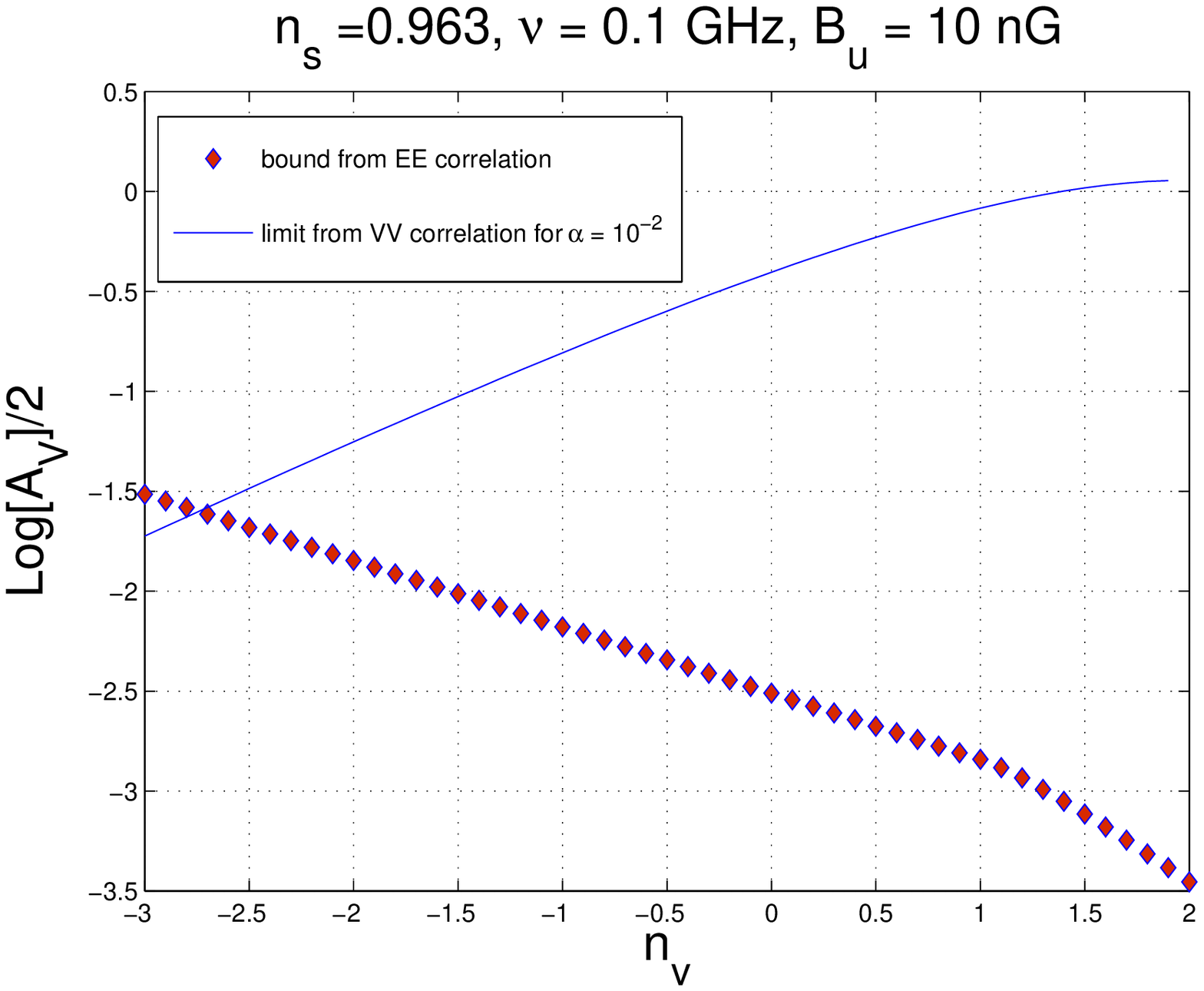}
\includegraphics[height=6cm]{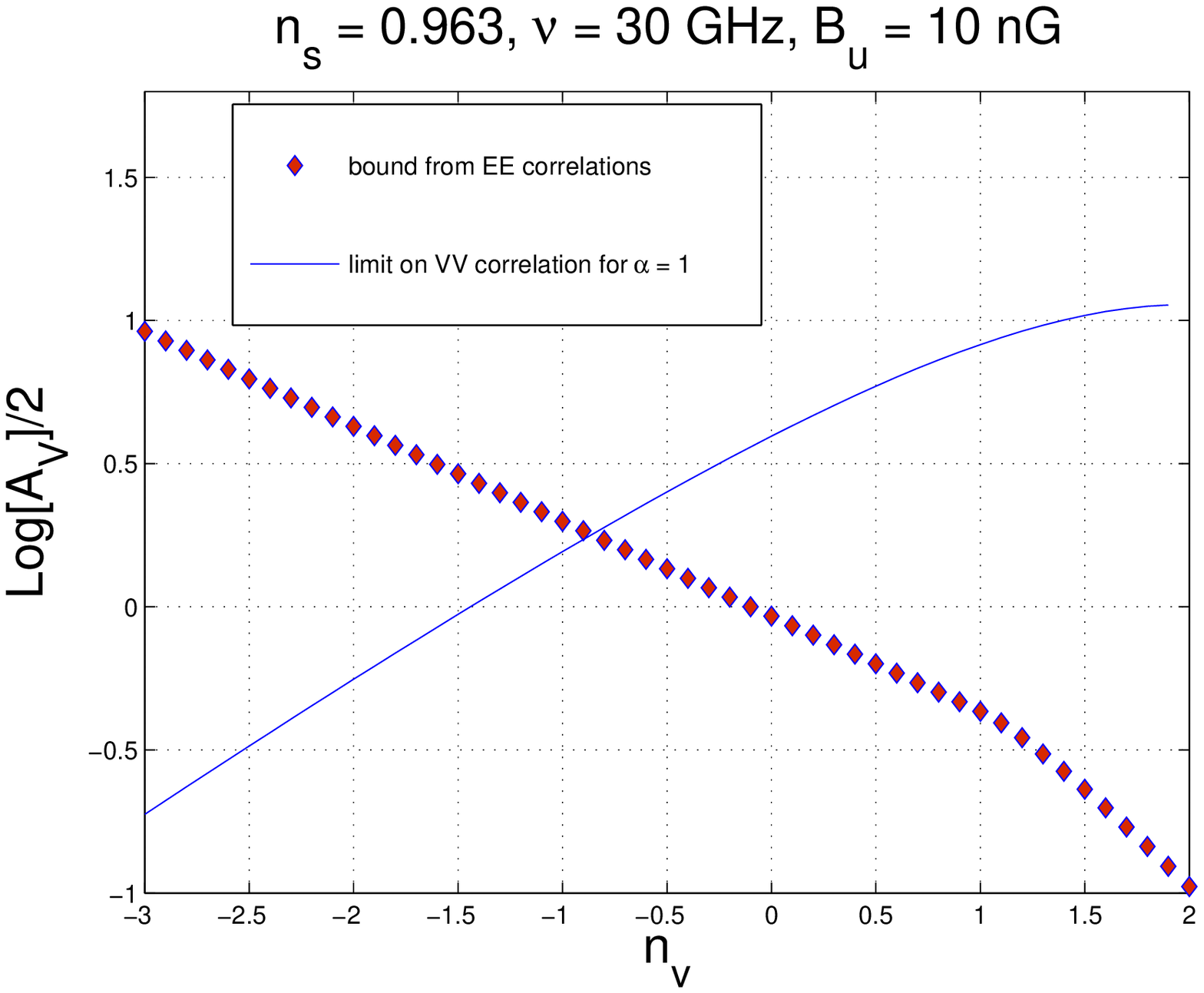}
\includegraphics[height=6cm]{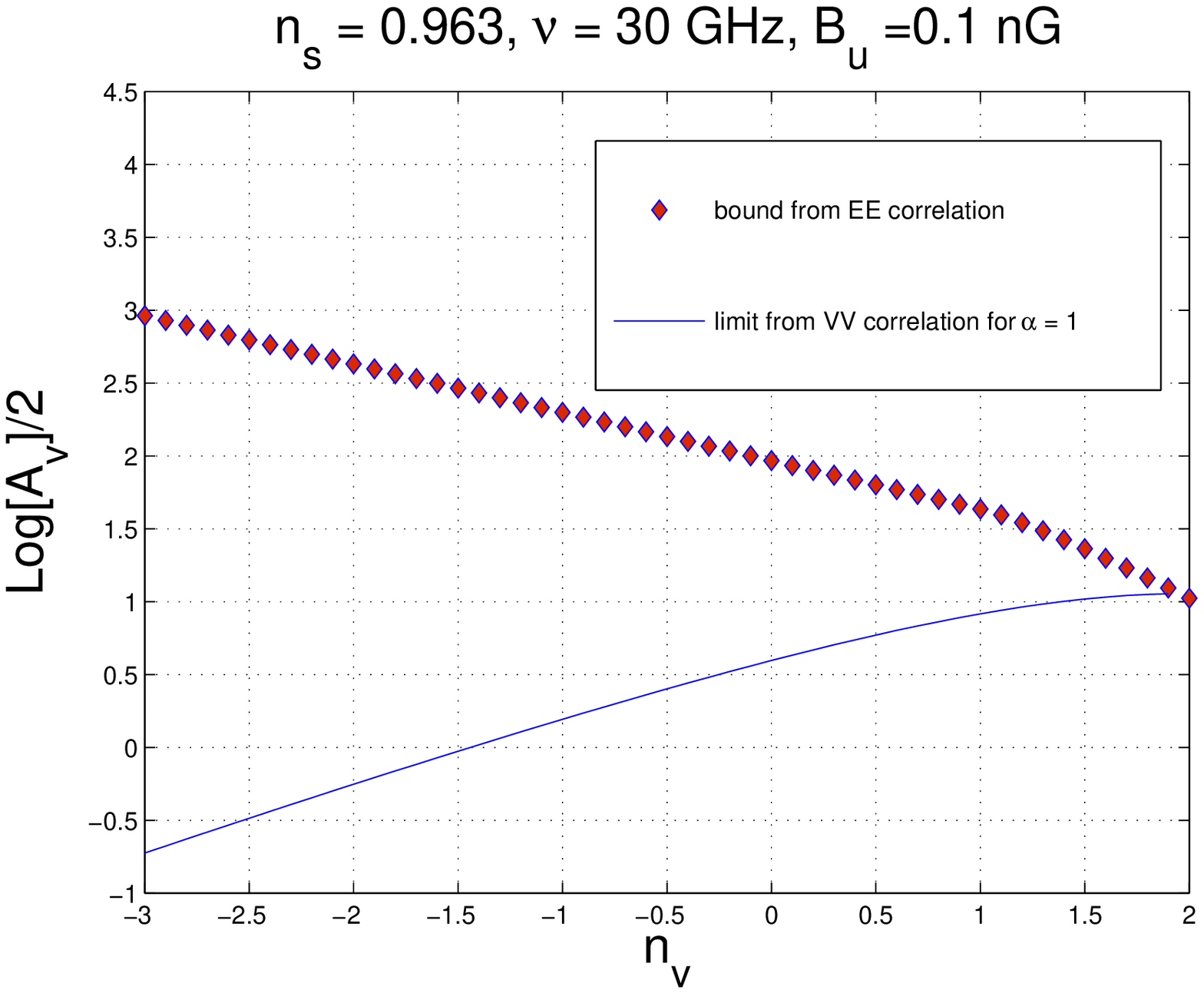}
\caption[a]{The bounds on the V-mode power spectrum as they arise 
from the EE correlations. The bounds derived in section \ref{sec4} are illustrated and compared 
with the potentially direct bounds parametrized, as in Fig. \ref{figure4}, in  terms of 
different values of $\alpha$ (see Eq. (\ref{limit8}).}
\label{figure5}      
\end{figure}
In Fig. \ref{figure4} with the full line we report the limit on 
$\sqrt{{\mathcal A}_{\mathrm{V}}}$ stemming from Eq. (\ref{limit8}) 
in terms of the corresponding value of $\alpha$. The various curves are 
 obtained by using, at the left hand side of Eq. (\ref{limit8}) 
the expression of Eq. (\ref{VM6}) appropriately averaged over the 
multipole range. Always in Fig. \ref{figure4} the starred points 
correspond to the bound on $\sqrt{{\mathcal A}_{\mathrm{V}}}$ 
derived in Eqs. (\ref{boundTT13}) and (\ref{boundTT14}). 

The results of Fig. \ref{figure4} suggest that 
for sufficiently large frequencies and for sufficiently small magnetic field 
intensity the bounds derived from the TT correlations are 
not competitive with direct limits. This aspect can  be appreciated from the 
two bottom plots of Fig. \ref{figure4} where already a value 
$\alpha =1 $ would imply a more stringent limit 
on $\sqrt{{\mathcal A}_{\mathrm{V}}}$.  Notice that the allowed region is below the
full line (if the limit of Eq. (\ref{limit8}) is considered) or below the 
starred points if the limit of Eqs. (\ref{boundTT13}) and (\ref{boundTT14}) is 
enforced. In connection with Fig. \ref{figure4} there are three possible situations:
\begin{itemize}
\item{} the full line could always be above the starred line: this never happens 
in the case of Fig. \ref{figure4} but it would simply mean that any indirect limit 
is more stringent than the direct one;
\item{} if the full line is below the starred line the indirect limit from the TT correlation is always compatible with the direct searches: this always happens if 
the magnetic field is sufficiently small (see, e.g. Fig. \ref{figure4} bottom right plot);
\item{} finally the full line may crosses the starred points: this is the most realistic 
situation in the light of the present and forthcoming direct limits on circular dichroism.
\end{itemize}
By looking at the top right and at the bottom left plots of Fig. \ref{figure4} 
it is then apparent that, depending upon the frequency of the experiment, 
magnetic fields $B_{\mathrm{u}}= {\mathcal O}(100\,\mathrm{nG})$ can be 
directly excluded for $\nu \simeq \mathrm{GHz}$ and with a sensitivity 
$\alpha \simeq 10^{-6}$ which would imply, in terms of Eq. (\ref{limit8}),
direct upper limits on the V-mode power spectrum ${\mathcal O}(\mu K)$ 
for $\ell < 40$. 

It should be stressed that  magnetic fields larger than the nG are now indirectly excluded by the joined analysis of the measured TT and TE power spectra \cite{mg3,mg4}. In this case a direct measurement of the V-mode polarization allows for independent constraints on the magnetic field intensity. At the same time 
the constraints derived in \cite{mg3,mg4} refer to fully inhomogeneous 
magnetic fields at large scales. Here we are constraining the uniform magnetic field 
affecting the scattering process. The field is clearly the same but its uniformity 
is just a consequence of the smallness of the photon wavelength in comparison 
with the inhomogeneity scale of the field. The inhomogeneity of the magnetic fields 
will also induce a Faraday rotation term which will add a coupling between the 
$\Delta_{\mathrm{Q}}$ and $\Delta_{\mathrm{U}}$. The Faraday  term will 
ultimately rotate the linear polarization of the CMB  \cite{far1,far2}. In this 
investigation possible contributions from Faraday mixing will be ignored
but their effects should be taken into account for more accurate estimates 
by following, for instance, the semi-analytic methodology introduced in 
\cite{far2}. 
 
In Fig. \ref{figure5} the same absolute bounds illustrated in Fig. \ref{figure4} are now compared with the bounds derived in section \ref{sec4} from the analysis of the EE correlations. The bounds stemming from the EE correlations are numerically more significant, especially for large spectral indices (i.e. $n_{v} >1$). As in the case of 
Fig. \ref{figure4} sufficiently small values of the magnetic field intensity 
make the indirect bounds rather loose in comparison with direct limits.
There are however numerical differences. 
From the top right and bottom left plots of Fig. \ref{figure5}
magnetic fields $B_{\mathrm{u}}= {\mathcal O}(10\,\mathrm{nG})$ can be 
directly excluded for $\nu \simeq \mathrm{GHz}$ and with a sensitivity 
$\alpha \simeq 10^{-6}$. The bounds stemming from the EE correlations 
are therefore more stringent than the ones derived in the case of the TT correlations. 

It is finally appropriate to mention that the frequency range assumed in the present 
discussion is in the GHz range because previous bounds, even if loose, were 
set over those frequencies. It is however tempting to speculate that, in a far future, 
CMB measurements could be possible even below the GHz. In this case 
direct bounds will certainly  be more stringent but huge foregrounds 
might make this speculation forlorn (see, in this connection, \cite{S1,S2}). 

\renewcommand{\theequation}{6.\arabic{equation}}
\setcounter{equation}{0}
\section{Concluding remarks}
\label{sec6}
A primordial degree of circular dichroism, uncorrelated with the standard adiabatic mode,
can be constrained if, prior to photon decoupling, the plasma is magnetized. 
The obtained results suggest the following considerations:
\begin{itemize}
\item{} if a V-mode power spectrum (not correlated with the adiabatic mode) 
is present prior to matter-radiation equality both the TT and the EE 
power spectra are affected in a computable manner;
\item{} constraints can then be inferred on the amplitude and spectral 
index of the V-mode power spectrum;
\item{} improved direct experimental limits on the VV 
correlations could be used for setting a limit on the magnetic field intensity.
\end{itemize}
For experimental devices operating in the 
GHz range, direct limits on the circular dichroism imply constraints  
on pre-decoupling magnetic fields in the $10$ nG range. Conversely, the current limits 
on large-scale magnetic fields derived from the distortions of the TT and TE 
correlations (in the $0.1$ nG range) are compatible with current bounds on
the primordial dichroism.  Improved bounds 
on the V-mode polarization are not only interesting in their own right but they
might have rewarding phenomenological implications. Direct limits on the V-mode power spectrum 
in the range ${\mathcal O}(0.01\, \mathrm{mK})$ imply limits on  ${\mathcal A}_{\mathrm{V}}$
ranging from  ${\mathcal O}(10^{-8})$ to  ${\mathcal O}(10^{-4})$ depending on the value of 
the spectral index and for large angular scales, i.e. larger than ${\mathcal O}(1\, \mathrm{deg})$.

\newpage
\begin{appendix}
\renewcommand{\theequation}{A.\arabic{equation}}
\setcounter{equation}{0}
\section{Explicit calculation of some angular integral}
\label{APPA}
In the explicit evaluation of various correlation functions it is often required to compute 
integrals involving the product of spherical harmonics and of various powers of $\mu$. Instead 
of computing all these integrals one by one we shall just refer to the general 
technique employed here. Consider the class of integrals of the type 
\begin{equation}
\int \, d \hat{n} f(\mu) \, Y_{\ell\, m}^{*}(\hat{n}) \, e^{- i \mu x}  
\label{AA1}
\end{equation}
where $f(\mu)$ is, for practical purposes, a polynomial in $\mu$ and where, as usual,
$d \hat{n} = d\mu \, d\varphi$. The integrals of the type (\ref{AA1}) can be evaluated by noticing that 
$f(\mu)$ can be replaced by $f(i \partial_{x})$ where $\partial_{x}$ denotes a derivation with respect 
to $x$. Alternatively it is possible to expand the exponential in Rayleigh series 
so that 
\begin{equation}
\int \, d \hat{n} f(\mu) \, Y_{\ell\, m}^{*}(\hat{n})   \, e^{- i \mu x}  = \sum_{j=0}^{\infty} (-i)^{j} (2j +1) \, j_{j}(x) {\mathcal U}_{\ell\, m}^{j},
\label{AA2}
\end{equation}
where the term ${\mathcal U}_{\ell\, m}^{j}$ is 
\begin{equation}
{\mathcal U}_{\ell\, m}^{j} = \int_{-1}^{1} d\mu \, \int_{0}^{2\pi} d\varphi \, f(\mu) P_{j}(\mu)  Y_{\ell\, m}^{*}(\mu,\varphi).
\label{AA3}
\end{equation}
If $f(\mu)$ is a polynomial in $\mu$ the integrand of Eq. (\ref{AA3}) can be simplified by performing 
directly the integration over $\varphi$ and by using the well known recurrence relation 
for the Legendre polynomials:
\begin{equation}
(j + 1) P_{j}(\mu) = ( 2 j + 1) \, \mu\, P_{j}(\mu) - j \, P_{j -1}(\mu).
\label{AA4}
\end{equation}
 Consider, for instance, the case where $f(\mu) = 1 + \mu^2$. In this case, using Eq. (\ref{AA4}), 
 Eq. (\ref{AA3}) becomes
\begin{equation}
{\mathcal U}_{\ell\, m}^{j} = \sqrt{4\pi ( 2 \ell + 1)} \, \delta_{m 0} \biggl[ \overline{A}(\ell) \delta_{\ell\, j}  + 
\overline{B}(\ell)  \delta_{j\, (\ell-2)} + \overline{C}(\ell) \delta_{j\, (\ell-2)}\biggr],
\label{AA5}
\end{equation}
where $\delta_{a\,b}$ is the Kroeneker delta function  and where
\begin{eqnarray}
&& \overline{A}(\ell) = \frac{(2\ell + 3) ( 5\ell^2-1) + (\ell+1)^2 (2\ell-1)}{(2 \ell + 1)^2 (2 \ell -1) ( 2 \ell +3)}, 
\nonumber\\
&& \overline{B}(\ell) = \frac{\ell (\ell -1)}{(2 \ell -1) ( 2 \ell +1)(2\ell -3)},\qquad 
\overline{C}(\ell) = \frac{(\ell +1) (\ell +2)}{(2 \ell + 3) ( 2 \ell +1)(2 \ell +5)}.
\label{AA6}
\end{eqnarray}
Inserting Eq. (\ref{AA6}) into Eq. (\ref{AA2}) 
\begin{equation}
\int \, d \hat{n} f(\mu) \, Y_{\ell\, m}^{*}(\hat{n})   \, e^{- i \mu x}  =
\sqrt{4\pi ( 2 \ell + 1)} \, \delta_{m 0} (-i)^{-\ell} \biggl[ A(\ell) j_{\ell}(x) - B(\ell) j_{\ell -2}(x) - C(\ell) j_{\ell + 2}(x) \biggr], 
\label{AA7}
\end{equation}
where 
\begin{equation}
 A(\ell) = ( 2 \ell +1) \overline{A}(\ell),\qquad B(\ell) = (2\ell -3) \overline{B}(\ell),\qquad 
 C(\ell) = (2\ell +5) \overline{C}(\ell).
\label{AA8}
\end{equation}
\end{appendix}

\end{document}